\documentclass[pdflatex,sn-mathphys-num]{sn-jnl}

\usepackage{graphicx}%
\usepackage{multirow}%
\usepackage{amsmath,amssymb,amsfonts}%
\usepackage{amsthm}%
\usepackage{mathrsfs}%
\usepackage[title]{appendix}%
\usepackage{xcolor}%
\usepackage{textcomp}%
\usepackage{manyfoot}%
\usepackage{booktabs}%
\usepackage{algorithm}%
\usepackage{algorithmicx}%
\usepackage{algpseudocode}%
\usepackage{listings}%
\usepackage{float}     
\usepackage{placeins}  
\usepackage{hyperref}

\theoremstyle{thmstyleone}%
\theoremstyle{thmstyletwo}%

\theoremstyle{thmstylethree}%

\begin{document}

\title[MCP Bridge]{MCP Bridge: A Lightweight, LLM-Agnostic RESTful Proxy for Model Context Protocol Servers}

\author[1]{\fnm{Arash} \sur{Ahmadi}}\email{arash.ahmadi-1@ou.edu}

\author[1]{\fnm{Sarah} \sur{Sharif}}\email{s.sh@ou.edu}

\author*[1]{\fnm{Yaser M.} \sur{Banad}}\email{bana@ou.edu}

\affil*[1]{\orgdiv{School of Electrical, and Computer Engineering}, \orgname{University of Oklahoma}, \orgaddress{\city{Norman}, \state{Oklahoma}, \country{United States}}}

\abstract{Large Language Models (LLMs) are increasingly augmented with external tools through standardized interfaces like the Model Context Protocol (MCP). However, current MCP implementations face critical limitations: they typically require local process execution through STDIO transports, making them impractical for resource-constrained environments like mobile devices, web browsers, and edge computing. We present MCP Bridge, a lightweight RESTful proxy that connects to multiple MCP servers and exposes their capabilities through a unified API. Unlike existing solutions, MCP Bridge is fully LLM-agnostic, supporting any backend regardless of vendor. The system implements a risk-based execution model with three security levels—standard execution, confirmation workflow, and Docker isolation—while maintaining backward compatibility with standard MCP clients. However, reliable execution within this framework requires models that can strictly adhere to protocol schemas. To this end, we also fine-tuned the Qwen3 4B and 8B model family on the Agent-Ark/Toucan-1.5M dataset using four Reinforcement Learning techniques: Group Relative Policy Optimization (GRPO), Dr. GRPO, Beta Normalization Policy Optimization (BNPO), and Decoupled Clip and Dynamic sAmpling Policy Optimization (DAPO). Evaluated on the MCPToolBench++ benchmark, our optimized model achieves an F1 score of 73.0\% that outperforms GPT-OSS-120B (62.17\%) and remains competitive with the 70B+ parameter baselines. Evaluation demonstrates that MCP Bridge successfully addresses the constraints of direct MCP connections while providing enhanced security controls and cross-platform compatibility, enabling sophisticated LLM-powered applications in previously inaccessible environments.}




\maketitle

\section{Introduction}\label{sec1}

Large Language Models (LLMs) have revolutionized natural language processing. They enable sophisticated conversational agents that can understand and generate human-like text across numerous domains \cite{Mialon2023}. Despite their impressive capabilities, these models are inherently limited by their training data and lack access to real-time information, specialized tools, and the ability to perform actions in external systems \cite{Qin2023}. To overcome these limitations, there has been a significant push toward augmenting LLMs with external tools and data sources, allowing them to retrieve information, execute computations, and interact with various services \cite{Lewis2020}.

The Model Context Protocol (MCP) represents a significant advancement in this direction, providing a standardized interface for connecting AI assistants to external tools and data sources \cite{Anthropic2024}. Introduced as an open protocol, MCP aims to establish a universal adapter—a ``USB-C port for AI applications''—that enables any compliant model to access any data repository or service through a consistent format. This standardization addresses the fragmentation problem where each new tool integration requires custom development, replacing it with a single, extensible protocol.

However, current MCP implementations face critical limitations that hinder widespread adoption. Many MCP servers rely on STDIO transports that require local process execution, making them impractical for resource-constrained environments such as edge devices, mobile applications, and web browsers. Direct connections to MCP servers from multiple isolated clients also create redundancy and increase resource usage, while the technical complexity of MCP tool formats poses barriers for non-expert users.

\begin{figure}[H]
\centering
\includegraphics[width=0.8\textwidth]{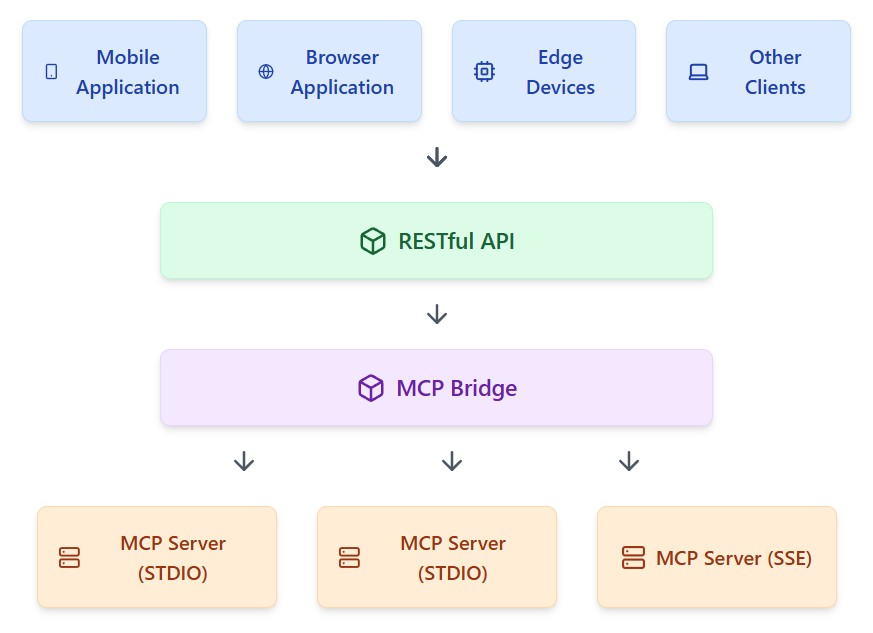}
\caption{Architecture of the MCP Bridge API system showing four layers: client applications (mobile, browser, edge devices, and others) at the top, connecting through a RESTful API to the MCP Bridge proxy, which interfaces with multiple MCP servers (STDIO and SSE) at the bottom. The system enables resource-constrained environments to access MCP functionality through a unified interface with configurable security levels.}\label{fig:architecture}
\end{figure}

In response to these challenges, we present MCP Bridge—a lightweight, fast, and LLM-agnostic proxy that connects to multiple MCP servers and exposes their capabilities through a unified REST API. The architecture is shown in Figure~\ref{fig:architecture}. While Anthropic's MCP SDK provides a reference client/server implementation, MCP Bridge focuses on deployment: it acts as a stable REST adapter that allows heterogeneous clients (e.g., browsers, mobile applications, edge devices) to use MCP servers without local process execution via STDIO transports. MCP Bridge also implements a risk-based execution model that supports standard execution, a confirmation workflow for medium-risk tools, and Docker isolation for high-risk tools, while maintaining backward compatibility with standard MCP clients. The implementation is available as an open-source project at \url{https://github.com/INQUIRELAB/mcp-bridge-api}.

\FloatBarrier

Although MCP Bridge is model-agnostic, end-to-end reliability depends on whether the client model can produce MCP-compliant tool calls consistently. To enable open-weight models to act as drop-in MCP Bridge clients, we fine-tune Qwen3-4B and Qwen3-8B using reinforcement learning on Toucan-1.5M \cite{xu2025toucan}, comparing GRPO \cite{shao2024deepseekmath}, Dr.\ GRPO \cite{liu2503understanding}, DAPO \cite{yu2025dapo}, and BNPO \cite{xiao2025bnpo}. We evaluate the resulting policies on MCPToolBench++ \cite{fan2025mcptoolbench++} and analyze the role of reward components via ablations (Appendix~\ref{app:reward-ablation}).

The remainder of this paper is organized as follows: Section~\ref{sec2} reviews related work in tool-augmented language models and standardization; Section~\ref{sec3} presents the MCP Bridge system design and implementation; Section~\ref{sec4} describes our policy optimization methodology for MCP tool alignment; Section~\ref{sec5} reports experimental results and comparisons; Section~\ref{sec6} discusses future directions; and Section~\ref{sec7} concludes. Reward ablations are provided in Appendix~\ref{app:reward-ablation}.

\section{Related Work}\label{sec2}

\subsection{Tool Use and Retrieval-Augmented Language Models}\label{subsec2.1}

Large language models (LLMs) have increasingly been augmented with external data sources and tools to overcome their inherent knowledge and capability limitations \cite{Mialon2023,Qin2023}. One prominent approach is retrieval-augmented generation (RAG), which integrates a document retriever with the model. Lewis et al. \cite{Lewis2020} introduced RAG as a general framework combining a parametric neural generator with non-parametric memory of retrieved documents, demonstrating improved performance on knowledge-intensive tasks. By linking to live knowledge sources, retrieval-augmented models can dynamically update their context and provenance, addressing issues like stale knowledge and hallucinations more effectively than static LLMs \cite{Lewis2020}.

Another research direction focuses on enabling LLMs to invoke external tools or APIs. Yao et al.'s ReAct framework \cite{Yao2023} interleaves logical reasoning traces with action commands, allowing models to make step-by-step decisions about when to continue thinking or call a tool. Schick et al. extended this concept with Toolformer \cite{Schick2023}, showing that language models can be self-taught to use tools through training on curated corpora with inserted API calls. HuggingGPT \cite{Shen2023} demonstrated model orchestration by using a powerful LLM as a controller that routes user requests to specialized AI models available.

The integration of tools with LLMs has been further advanced through frameworks like Visual ChatGPT \cite{Wu2023}, which connects LLMs with visual foundation models, and GPT-4Tools \cite{Yang2023}, which employs self-instruction to teach models new tool-use skills. Lu et al. developed Chameleon \cite{Lu2023}, a plug-and-play reasoning framework that augments LLMs with modular tools and uses an LLM-based planner to coordinate tool composition. Gorilla \cite{Patil2023} addresses real-world API invocation by fine-tuning LLaMA to output exact API calls from a large catalog of machine learning APIs.

\subsection{Standardization and LLM-Agnostic Integration}\label{subsec2.2}

As the ecosystem of LLM-accessible tools expands, integration scalability has emerged as a key challenge—specifically, how to connect any model to any tool with minimal custom code. Anthropic's Model Context Protocol (MCP) \cite{Anthropic2024} represents a significant standardization effort, introduced as an open standard for connecting AI assistants to software and data. MCP provides a unified client-server architecture where tools and data sources are wrapped as MCP servers that expose specific functions through a consistent interface.

In parallel to protocol standardization, researchers have explored RESTful API proxies as bridges between LLMs and existing web services. RestGPT by Song et al. \cite{Song2023} treats REST APIs as tools that an LLM can learn to use by providing the model with OpenAPI specifications and implementing a coarse-to-fine planning approach. It introduces a dedicated API executor that handles HTTP requests and response parsing, enabling a single framework to access hundreds of different APIs through a uniform method.

Beyond research prototypes, practical frameworks like LangChain and GPT-Index provide unified APIs to access multiple LLM backends and incorporate external tools, reflecting the need for abstraction layers. However, these solutions are largely engineering-focused, whereas academic efforts like MCP aim to establish formal standards that encourage compatibility across diverse systems and vendors.

The MCP Bridge proposed in this work follows this standardization philosophy by acting as a RESTful adapter between MCP servers and client applications. By building on prior ideas of tool augmentation and standard interfaces, MCP Bridge delivers a practical, modular integration solution that remains agnostic to the underlying model or environment. This approach aligns with the broader movement toward making advanced AI functionalities more accessible, interoperable, and future-proof for real-world applications.

\section{System Design and Implementation}\label{sec3}

This section describes the design and implementation of MCP Bridge, a lightweight, fast, and LLM-agnostic proxy for Model Context Protocol (MCP) servers. We detail the system architecture, API design, server management, security model, and client integration components.

\subsection{System Architecture and Technology Stack}\label{subsec3.1}

MCP Bridge follows a layered architecture that decouples client applications from the underlying MCP server processes. Figure~\ref{fig:architecture} illustrates this design, where client applications communicate with the proxy via a standardized REST API, and the proxy manages connections to multiple MCP servers.

The system is built on Node.js (18+) and uses the following core components:
\begin{itemize}
    \item \textbf{Express.js}: Provides the HTTP server and routing capabilities
    \item \textbf{Child Process API}: Manages spawned MCP server processes
    \item \textbf{Server-Sent Events (SSE)}: Enables real-time communication between some MCP servers and the proxy
    \item \textbf{Docker SDK}: Facilitates containerized execution for high-risk operations
\end{itemize}

This technology stack was chosen for its minimal footprint, cross-platform compatibility, and non-blocking I/O capabilities—critical requirements for a proxy that must handle multiple concurrent connections with low latency. The implementation uses asynchronous programming patterns throughout to prevent blocking operations from degrading performance.

\subsection{RESTful API and Endpoints}\label{subsec3.2}

MCP Bridge exposes a comprehensive REST API that standardizes access to MCP server functionality. The API is organized into general endpoints for server management and server-specific endpoints for tool execution and resource access.

Table~\ref{tab:endpoints} summarizes the primary API endpoints provided by MCP Bridge.

\begin{table}[h]
\caption{MCP Bridge API Endpoints}\label{tab:endpoints}
\begin{tabular}{@{}lll@{}}
\toprule
Endpoint & Method & Description \\
\midrule
\texttt{/servers} & GET & List all connected MCP servers \\
\texttt{/servers} & POST & Start a new MCP server \\
\texttt{/servers/\{serverId\}} & DELETE & Stop and remove a server \\
\texttt{/health} & GET & Get health status of MCP Bridge \\
\texttt{/confirmations/\{id\}} & POST & Confirm execution of a medium-risk request \\
\texttt{/servers/\{id\}/tools} & GET & List all tools for a specific server \\
\texttt{/servers/\{id\}/tools/\{toolName\}} & POST & Execute a specific tool \\
\texttt{/servers/\{id\}/resources} & GET & List all resources \\
\texttt{/servers/\{id\}/prompts} & GET & List all prompts \\
\bottomrule
\end{tabular}
\end{table}

The API design follows REST principles with JSON as the primary data exchange format. Each endpoint returns appropriate HTTP status codes and standardized error responses. For example, when executing a tool via \texttt{POST /servers/\{id\}/tools/\{toolName\}}, the request body contains the tool's input parameters, and the response includes the execution result or a confirmation request based on the tool's risk level.

\begin{algorithm}
\caption{API Request Processing Pipeline}\label{alg:request-processing}
\begin{algorithmic}[1]
\State \textbf{Input:} HTTP request $req$ with server ID $sid$, tool name $tool$, and parameters $params$
\State \textbf{Output:} HTTP response $res$ with result or confirmation request

\Function{ProcessToolRequest}{$req, res$}
    \State $sid \gets req.params.serverId$
    \State $tool \gets req.params.toolName$
    \State $params \gets req.body$
    
    \If{not \Call{ServerExists}{$sid$}}
        \State \Return \Call{Error}{$res, 404, \text{"Server not found"}$}
    \EndIf
    
    \State $server \gets \Call{GetServer}{sid}$
    \If{not \Call{ToolExists}{$server, tool$}}
        \State \Return \Call{Error}{$res, 404, \text{"Tool not found"}$}
    \EndIf
    
    \State $riskLevel \gets \Call{GetToolRiskLevel}{server, tool}$
    
    \If{$riskLevel = 1$}
        \State $result \gets \Call{ExecuteTool}{server, tool, params}$
        \State \Return \Call{Success}{$res, result$}
    \ElsIf{$riskLevel = 2$}
        \State $confirmationId \gets \Call{GenerateConfirmationId}{}$
        \State \Call{StoreConfirmationRequest}{$confirmationId, server, tool, params$}
        \State \Return \Call{RequireConfirmation}{$res, confirmationId$}
    \ElsIf{$riskLevel = 3$}
        \State $result \gets \Call{ExecuteToolInDocker}{server, tool, params}$
        \State \Return \Call{Success}{$res, result$}
    \EndIf
\EndFunction
\end{algorithmic}
\end{algorithm}

The request processing pipeline (Algorithm~\ref{alg:request-processing}) shows how MCP Bridge handles tool execution requests, including validation, risk assessment, and appropriate execution pathways. This unified API layer provides consistent access patterns regardless of the underlying MCP server implementation.

\subsection{Server Management and Connection Handling}\label{subsec3.3}

MCP Bridge dynamically manages connections to MCP servers, supporting both standard STDIO-based servers and newer Server-Sent Events (SSE) implementations. The server management subsystem handles server lifecycle (startup, monitoring, and teardown) and efficiently routes requests to the appropriate server instance.
\begin{algorithm}
\caption{MCP Server Connection Management}
\label{alg:server-management}
\begin{algorithmic}[1]
\State \textbf{Input:} Server configuration $config$ with command, arguments, and environment variables
\State \textbf{Output:} Server connection object or error

\Function{StartMcpServer}{$config$}
    \State $serverId \gets \Call{GenerateUUID}{}$
    \State $process \gets \text{null}$
    
    \Comment{Attempt to start MCP server}
    \If{$config.command$ exists}
        \State $process \gets \Call{SpawnProcess}{config.command, config.args, config.env}$
    \Else
        \State \Return $\text{Error: "Invalid server configuration"}$
    \EndIf
    
    \State $connected \gets \Call{WaitForConnection}{process, timeout=5000}$
    \If{not $connected$}
        \State \Call{KillProcess}{process}
        \State \Return $\text{Error: "Failed to connect to MCP server"}$
    \EndIf
    
    \State $serverConn \gets \{id: serverId, process: process, tools: [], resources: []\}$
    \State \Call{DiscoverServerCapabilities}{serverConn}
    \State \Call{RegisterServer}{serverId, serverConn}
    \State \Return $serverConn$
    
    \Comment{Error handling}
    \If{an error occurs}
        \If{$process \neq \text{null}$}
            \State \Call{KillProcess}{process}
        \EndIf
        \State \Return $\text{Error: error.message}$
    \EndIf
\EndFunction
\end{algorithmic}
\end{algorithm}

As shown in Algorithm~\ref{alg:server-management}, server connections are established by spawning child processes or connecting to existing MCP servers via their specified transport. The system automatically discovers each server's capabilities (available tools, resources, and prompts) upon connection, making them immediately available through the REST API.

The connection manager employs several strategies to maintain robust connections:

\begin{itemize}
    \item \textbf{Heartbeat monitoring}: Periodically checks server health
    \item \textbf{Automatic reconnection}: Attempts to re-establish lost connections
    \item \textbf{Connection pooling}: Optimizes resource usage for high-demand servers
    \item \textbf{Request queueing}: Manages concurrent requests to prevent overloading servers
\end{itemize}

These mechanisms ensure that client applications experience minimal disruption even when underlying MCP servers encounter issues or need to be restarted.

\subsection{Security Model and Risk-Based Execution}\label{subsec3.4}

MCP Bridge implements a comprehensive security model centered around risk-based execution levels. This approach provides granular control over tool invocation, particularly for operations that could potentially modify data or access sensitive resources.

The risk-based execution model defines three levels:

\begin{enumerate}
    \item \textbf{Low Risk (Level 1)}: Standard execution without additional checks, suitable for read-only operations
    \item \textbf{Medium Risk (Level 2)}: Requires explicit confirmation before execution, appropriate for data-modifying operations
    \item \textbf{High Risk (Level 3)}: Executed within an isolated Docker container, providing environmental isolation for maximum security
\end{enumerate}

The workflow for medium-risk operations is particularly important (see Algorithm~\ref{alg:confirmation-flow}), as it implements a two-phase execution pattern that requires explicit confirmation before proceeding.

\begin{algorithm}
\caption{Medium-Risk Confirmation Workflow}\label{alg:confirmation-flow}
\begin{algorithmic}[1]
\State \textbf{Input:} Confirmation ID $confirmationId$, confirmation token $token$
\State \textbf{Output:} Execution result or error

\Function{ProcessConfirmation}{$confirmationId, token$}
    \State $pendingReq \gets \Call{GetPendingRequest}{confirmationId}$
    \If{$pendingReq = \text{null}$}
        \State \Return $\text{Error: "Invalid confirmation ID or expired request"}$
    \EndIf
    
    \If{$pendingReq.token \neq token$}
        \State \Return $\text{Error: "Invalid confirmation token"}$
    \EndIf
    
    \If{$\Call{IsExpired}{pendingReq}$}
        \State \Call{RemovePendingRequest}{confirmationId}
        \State \Return $\text{Error: "Confirmation expired"}$
    \EndIf
    
    \State $server \gets \Call{GetServer}{pendingReq.serverId}$
    \State $result \gets \Call{ExecuteTool}{server, pendingReq.tool, pendingReq.params}$
    \State \Call{RemovePendingRequest}{confirmationId}
    \State \Return $result$
\EndFunction
\end{algorithmic}
\end{algorithm}

For high-risk operations (Level 3), MCP Bridge leverages Docker containers to provide strong isolation. Each container is configured with specific resource limits, network controls, and volume mounts as defined in the server configuration. This containerization ensures that even if a tool behaves unexpectedly, its impact is contained within the isolated environment.

This multi-tiered approach to security allows system administrators to configure appropriate risk levels based on their security requirements while maintaining compatibility with standard MCP clients that expect direct execution.

\subsection{Client Integration (MCP-Gemini Agent)}\label{subsec3.5}

\begin{figure}[H]
\centering
\includegraphics[width=0.4\textwidth]{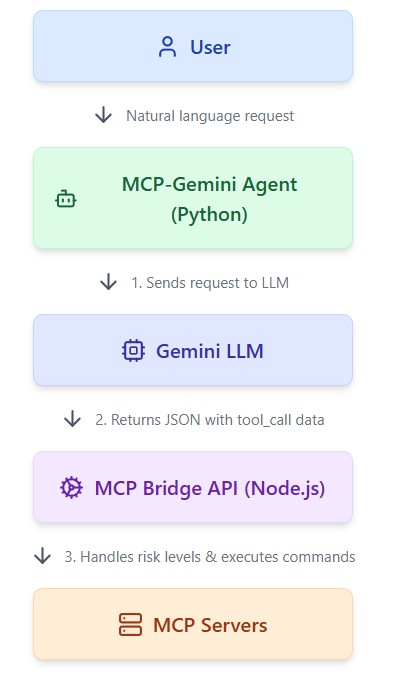}
\caption{This figure illustrates the integration between a large language model (Gemini) and the Model Context Protocol (MCP) Bridge. The architecture shows a vertical flow from the user through the MCP-Gemini Agent (Python), to the Gemini LLM, then to the MCP Bridge API (Node.js), and finally to the MCP Servers. The diagram highlights the system's key data flows: natural language inputs from users, structured tool calls from the LLM, and request execution with risk-level handling (low, medium, high). The system supports multi-step reasoning by cycling results back to the LLM to determine subsequent actions.}\label{fig:gemini}
\end{figure}

Complementing the server-side proxy is the MCP-Gemini Agent, a Python client that integrates Google's Gemini API with MCP Bridge. This agent provides an intelligent natural language interface to the MCP tool ecosystem, allowing users to interact with tools through conversational prompts rather than direct API calls. Figure \ref{fig:gemini} shows the structure of the LLM integration.

\FloatBarrier

The MCP-Gemini Agent implements several key features:

\begin{itemize}
    \item \textbf{Multi-step reasoning}: Supports complex operations by sequencing tool calls
    \item \textbf{Security confirmation handling}: Seamlessly manages the confirmation workflow for medium-risk operations
    \item \textbf{Flexible JSON display}: Configurable verbosity for tool outputs
    \item \textbf{Automatic tool discovery}: Detects and utilizes all available tools from connected servers
\end{itemize}

The agent's architecture follows a conversational loop pattern (see Algorithm~\ref{alg:gemini-agent}), where user inputs are processed by the Gemini LLM to generate appropriate tool calls to MCP Bridge.

\begin{algorithm}
\caption{MCP-Gemini Agent Conversation Loop}\label{alg:gemini-agent}
\begin{algorithmic}[1]
\State \textbf{Input:} User query $query$, MCP Bridge URL $mcpUrl$
\State \textbf{Output:} Agent response with tool execution results

\Function{ProcessUserQuery}{$query, mcpUrl$}
    \State $toolList \gets \Call{FetchAvailableTools}{mcpUrl}$
    \State $prompt \gets \Call{BuildToolAwarePrompt}{query, toolList}$
    \State $llmResponse \gets \Call{InvokeGeminiLLM}{prompt}$
    
    \State $tools \gets \Call{ExtractToolCalls}{llmResponse}$
    \State $results \gets$ empty list
    
    \For{each $tool$ in $tools$}
        \State $result \gets \Call{ExecuteMcpTool}{mcpUrl, tool.serverId, tool.name, tool.params}$
        
        \If{$result.requiresConfirmation$}
            \State $confirmation \gets \Call{PromptUserForConfirmation}{result.confirmationDetails}$
            \If{$confirmation.confirmed$}
                \State $result \gets \Call{ConfirmToolExecution}{mcpUrl, result.confirmationId}$
            \Else
                \State $result \gets \{status: "cancelled", message: "User cancelled operation"\}$
            \EndIf
        \EndIf
        
        \State Append $result$ to $results$
    \EndFor
    
    \State $followupPrompt \gets \Call{BuildResultPrompt}{query, tools, results}$
    \State $finalResponse \gets \Call{InvokeGeminiLLM}{followupPrompt}$
    \State \Return $finalResponse$
\EndFunction
\end{algorithmic}
\end{algorithm}

The MCP-Gemini Agent demonstrates that proprietary LLMs can use MCP Bridge without additional training by emitting structured tool calls directly; in Section~\ref{sec4}, we show that open-weight models benefit from explicit RL alignment to satisfy MCP’s strict formatting and tool-selection requirements.

The agent implementation is modular and configurable, supporting various command-line options for customizing behavior:

\begin{verbatim}
usage: llm_test.py [-h] [--hide-json] [--json-width JSON_WIDTH] 
                   [--mcp-url MCP_URL] [--mcp-port MCP_PORT]

MCP-Gemini Agent with configurable settings

options:
  -h, --help            show this help message and exit
  --hide-json           Hide JSON results from tool executions
  --json-width JSON_WIDTH
                        Maximum width for JSON output (default: 100)
  --mcp-url MCP_URL     MCP Bridge URL (default: http://localhost:3000)
  --mcp-port MCP_PORT   Override port in MCP Bridge URL
\end{verbatim}

Beyond demonstrating MCP Bridge with a proprietary LLM client, the proxy is designed to be \emph{model-agnostic}: any language model capable of emitting MCP-style structured tool calls can use the same REST endpoints to discover tools, invoke them, and receive results. In the next section, we present an empirical evaluation of MCP Bridge's system-level performance before describing how we fine-tune open-weight Qwen3 models for MCP tool alignment.

\subsection{System Performance Evaluation}\label{subsec3.6}

To validate the architectural claims presented above and quantify the performance characteristics of the REST proxy layer, we conducted a systematic benchmark of MCP Bridge measuring end-to-end latency, throughput under concurrent load, risk-level execution-path overhead, and resource utilization. All measurements were collected on a single machine running Windows~11 with an AMD Ryzen~7 7435HS (16 logical cores) and 24,261~MB RAM, using Node.js v22.12.0. To ensure statistical reliability, latency and concurrency tests were repeated over three independent runs with 50 iterations per operation per run; we report mean~$\pm$~standard deviation across runs.

\subsubsection{Benchmark Configuration}
Four MCP servers were started simultaneously through MCP Bridge using the configuration summarized in Table~\ref{tab:bench-env}. The servers span three distinct MCP server implementations and two risk levels, exercising 50 tools in total.

\begin{table}[h]
\centering
\caption{MCP Bridge benchmark configuration. Four STDIO-based MCP servers were started simultaneously, spanning three distinct server implementations and two risk levels.}
\label{tab:bench-env}
\begin{tabular}{@{}llcl@{}}
\toprule
Server ID & Implementation & Risk Level & Tools \\
\midrule
\texttt{filesystem}        & \texttt{@modelcontextprotocol/server-filesystem} & 1 (Low)    & 14 \\
\texttt{filesystem-medium} & \texttt{@modelcontextprotocol/server-filesystem} & 2 (Medium) & 14 \\
\texttt{memory}            & \texttt{@modelcontextprotocol/server-memory}     & 1 (Low)    & 9  \\
\texttt{everything}        & \texttt{@modelcontextprotocol/server-everything}  & 1 (Low)    & 13 \\
\midrule
\multicolumn{3}{l}{\textbf{Total tools discovered}} & \textbf{50} \\
\bottomrule
\end{tabular}
\end{table}
\FloatBarrier

\subsubsection{Latency Analysis}
We measured round-trip latency for six representative tool-execution operations under three access modes: (i) \emph{Bridge REST}, where the client sends an HTTP request to MCP Bridge and the proxy forwards it to the MCP server over STDIO; (ii) \emph{STDIO keep-alive}, a direct STDIO connection that remains open across requests (best-case baseline); and (iii) \emph{STDIO per-spawn}, where a new server process is spawned for every request (worst-case baseline). Table~\ref{tab:latency-comparison} reports the results.

\begin{table}[h]
\centering
\caption{Tool-execution latency comparison (ms). Bridge REST values are mean~$\pm$~std across 3 runs of 50 iterations each. STDIO values are from a single run (50 iterations keep-alive, 10 iterations per-spawn). Bridge overhead is Bridge REST mean minus STDIO keep-alive mean.}
\label{tab:latency-comparison}
\begin{tabular}{@{}lcccc@{}}
\toprule
Operation & Bridge REST & STDIO Keep-Alive & STDIO Per-Spawn & Overhead \\
\midrule
\texttt{filesystem/list\_directory}  & 2.55 $\pm$ 0.35 & 0.91 & 5.60 & 1.64 \\
\texttt{filesystem/read\_file}       & 2.44 $\pm$ 0.03 & 1.10 & 6.72 & 1.34 \\
\texttt{everything/echo}             & 1.55 $\pm$ 0.10 & 0.48 & 5.24 & 1.07 \\
\texttt{everything/get-sum}          & 1.66 $\pm$ 0.08 & ---  & ---  & ---  \\
\texttt{memory/create\_entities}     & 2.98 $\pm$ 0.08 & ---  & ---  & ---  \\
\texttt{memory/read\_graph}          & 2.15 $\pm$ 0.08 & 0.80 & 5.03 & 1.35 \\
\bottomrule
\end{tabular}
\end{table}
\FloatBarrier

\begin{figure}[H]
\centering
\includegraphics[width=0.85\textwidth]{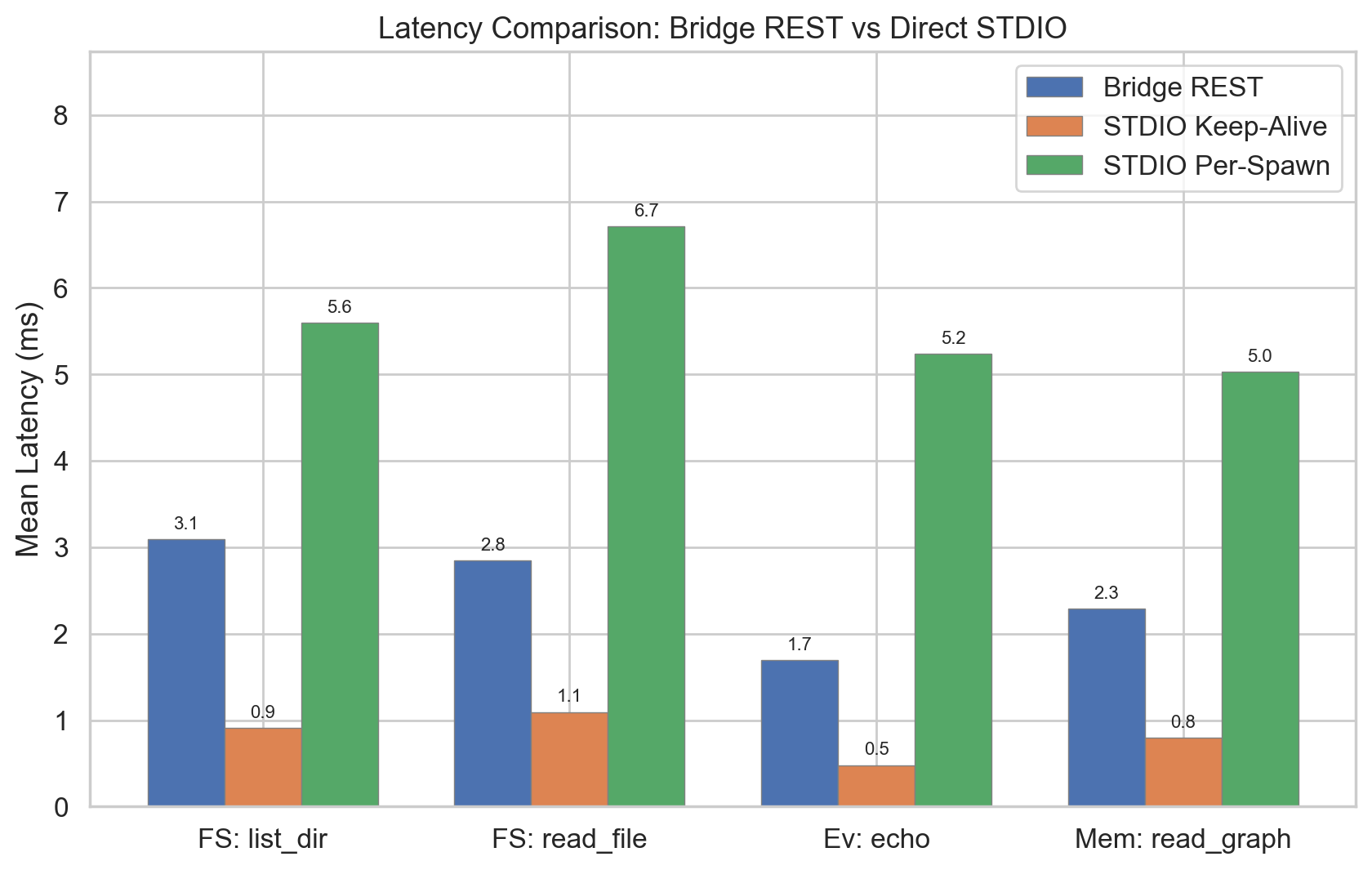}
\caption{Tool-execution latency comparison across three access modes: Bridge REST, direct STDIO keep-alive, and direct STDIO per-spawn. The REST proxy adds 1--2~ms over a persistent STDIO pipe but is 2.5--4.3$\times$ faster than spawning a new server process per request, which is the typical scenario for resource-constrained clients.}\label{fig:latency-comparison}
\end{figure}

The Bridge REST proxy adds a mean overhead of 1.07--1.64~ms over a persistent STDIO connection, attributable to HTTP parsing, JSON serialization, and Express.js routing. Critically, Bridge REST is \emph{faster} than per-spawn STDIO by 2.5--4.3$\times$ because the proxy maintains long-lived server processes; clients that would otherwise need to spawn a new MCP server for each request (the typical scenario for browsers or mobile apps) see a substantial latency reduction. Figure~\ref{fig:latency-comparison} visualizes the three-way comparison.

\FloatBarrier

\subsubsection{Throughput Under Concurrent Load}
\begin{figure}[H]
\centering
\includegraphics[width=0.85\textwidth]{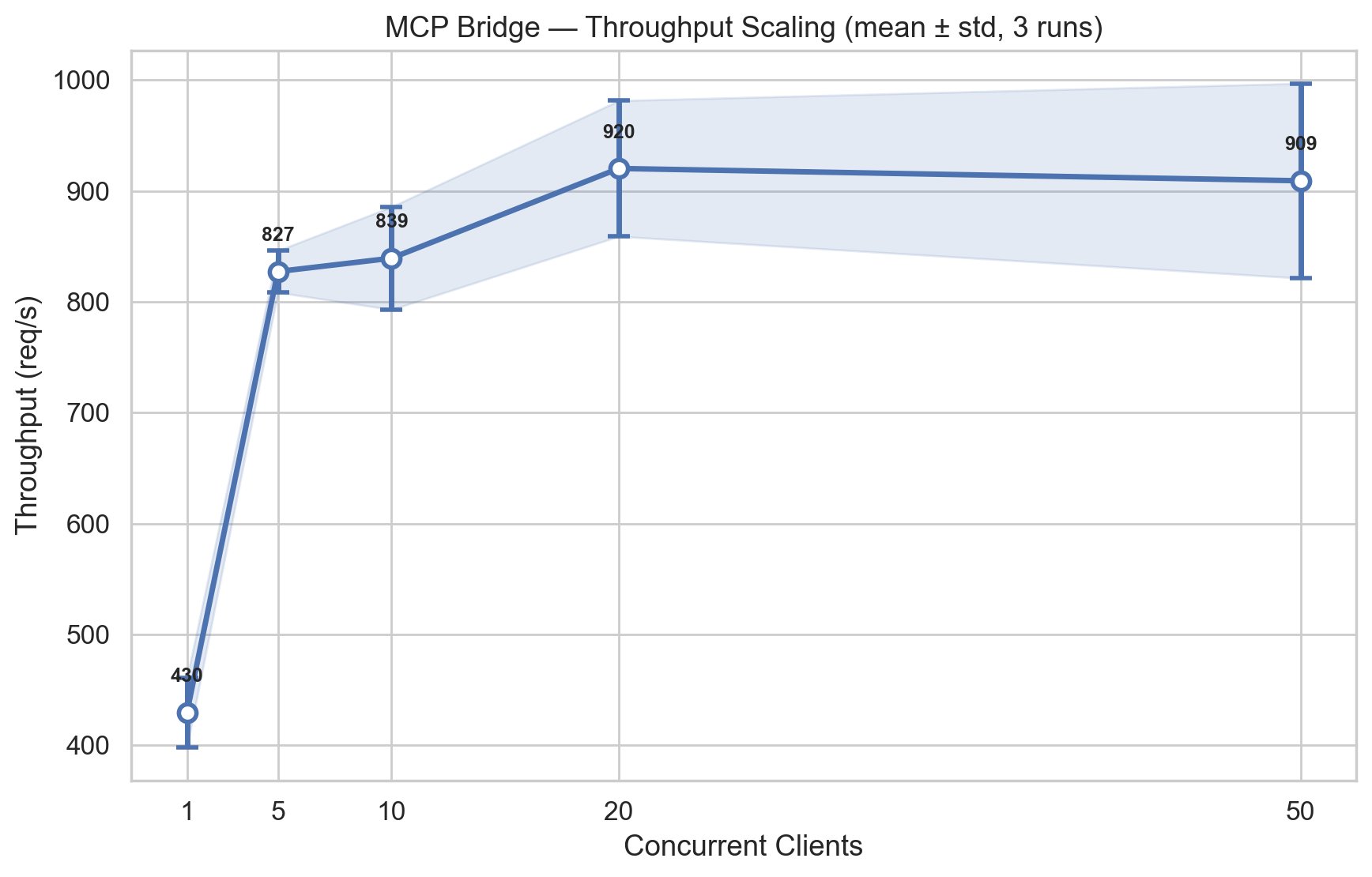}
\caption{Throughput scaling as a function of concurrent clients (mean~$\pm$~std, 3 independent runs). MCP Bridge doubles throughput from 1 to 5 clients and maintains $>$900~req/s at 50 concurrent clients with zero errors.}\label{fig:throughput-scaling}
\end{figure}

To evaluate scalability, we measured throughput (requests/second) and mean latency at concurrency levels of 1, 5, 10, 20, and 50 simultaneous clients. Each level issued 20--100 mixed tool-execution requests distributed across the four servers. Table~\ref{tab:concurrency} and Figure~\ref{fig:throughput-scaling} report the results.

\begin{table}[h]
\centering
\caption{Throughput and latency under concurrent load (mean~$\pm$~std, 3 runs). MCP Bridge maintains zero errors across all concurrency levels.}
\label{tab:concurrency}
\begin{tabular}{@{}ccccc@{}}
\toprule
Concurrency & Throughput (req/s) & Mean Latency (ms) & Total Requests & Errors \\
\midrule
1  & 430 $\pm$ 31  & 2.26 $\pm$ 0.15 & 20  & 0 \\
5  & 827 $\pm$ 19  & 5.08 $\pm$ 0.10 & 100 & 0 \\
10 & 839 $\pm$ 46  & 10.47 $\pm$ 0.53 & 100 & 0 \\
20 & 920 $\pm$ 61  & 18.20 $\pm$ 1.05 & 100 & 0 \\
50 & 909 $\pm$ 88  & 43.14 $\pm$ 3.63 & 100 & 0 \\
\bottomrule
\end{tabular}
\end{table}
\FloatBarrier

Throughput scales from 430~req/s (single client) to over 900~req/s at 50 concurrent clients, demonstrating that the asynchronous Node.js architecture effectively exploits I/O concurrency. Latency increases approximately linearly with concurrency, which is expected for a single-threaded event-loop proxy, while the error rate remains at 0\% across all levels.

\FloatBarrier

\subsubsection{Risk-Level Execution Traces}
To validate each execution path described in Section~\ref{subsec3.4}, we recorded end-to-end latency traces for the three risk levels using real MCP server requests.

\begin{itemize}
    \item \textbf{Level~1 (Direct)}: A \texttt{list\_directory} call to the \texttt{filesystem} server (risk level~1) completed in 4.61~ms with a single HTTP round trip.
    \item \textbf{Level~2 (Confirmation)}: The same tool on \texttt{filesystem-medium} (risk level~2) required two round trips: the initial request returned a confirmation token in 1.35~ms, and execution after confirmation completed in 6.82~ms, totaling 8.16~ms. Re-submitting the same token returned HTTP~404, confirming single-use enforcement.
    \item \textbf{Level~2 (Rejection)}: Submitting a rejection for a pending confirmation correctly returned a cancellation response without executing the tool.
\end{itemize}

\begin{figure}[H]
\centering
\includegraphics[width=0.75\textwidth]{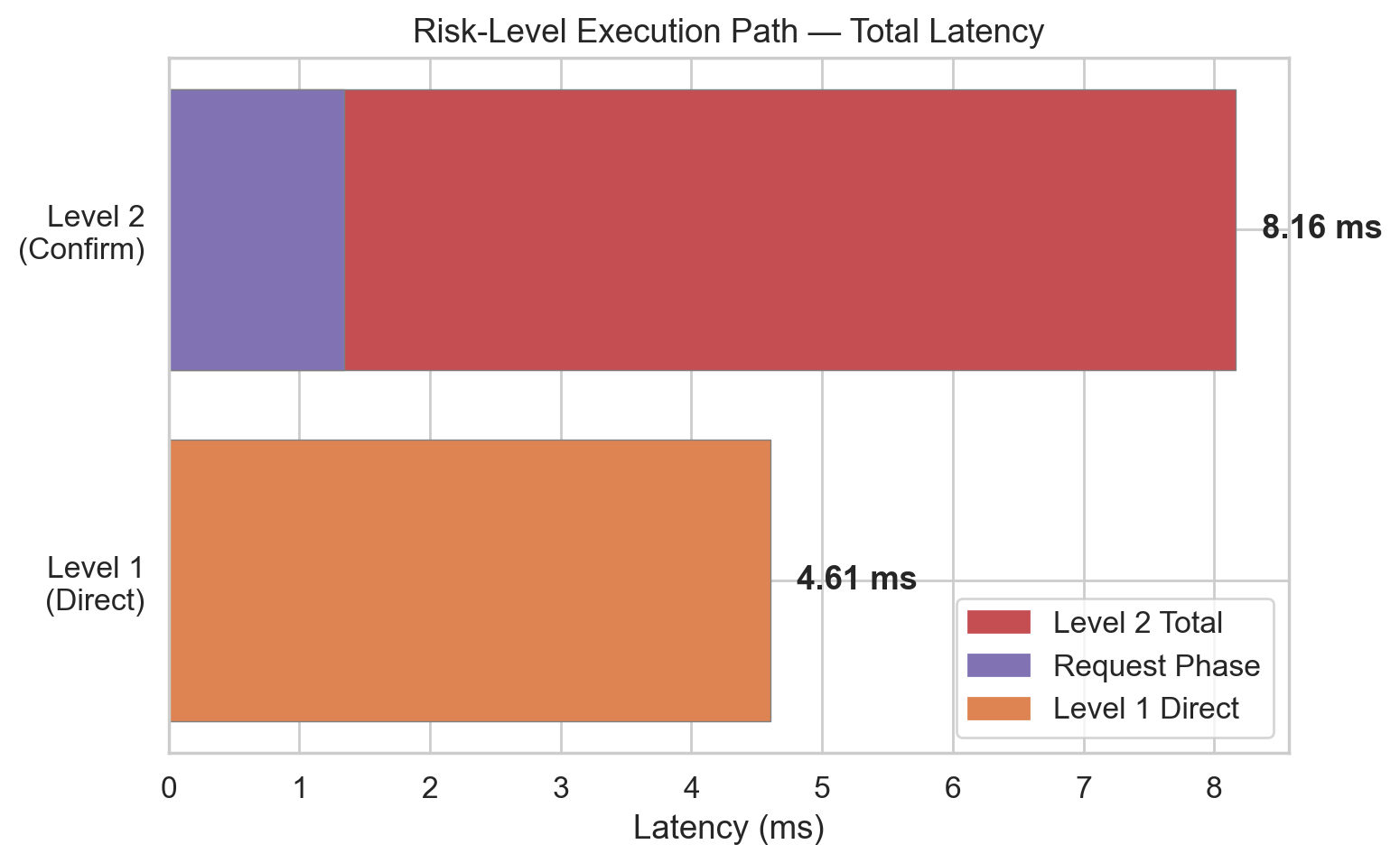}
\caption{Risk-level execution-path latency. Level~1 (direct) completes in a single round trip; Level~2 (confirmation) requires two round trips with a single-use token validator.}\label{fig:risk-level-traces}
\end{figure}

Figure~\ref{fig:risk-level-traces} shows the latency breakdown. The confirmation workflow adds approximately 3.6~ms of overhead compared to direct execution, a modest cost for the additional safety guarantee.

\FloatBarrier

\subsubsection{Resource Utilization and Cold Start}
\begin{figure}[H]
\centering
\includegraphics[width=0.85\textwidth]{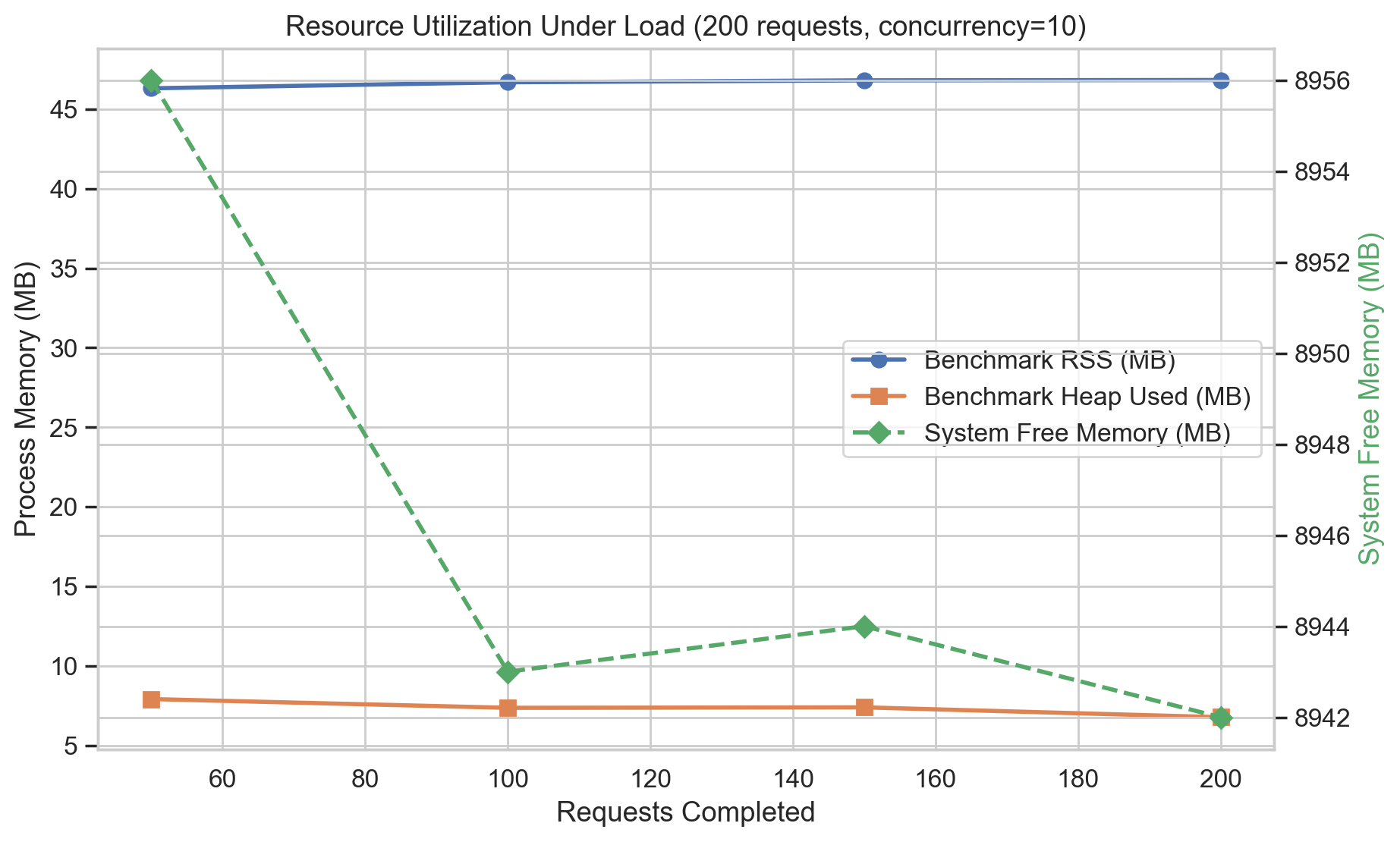}
\caption{Resource utilization during a sustained-load test (200 requests, concurrency~10). Process RSS and heap usage remain stable, and system free memory is unaffected, confirming a minimal resource footprint under load.}\label{fig:resource-utilization}
\end{figure}

Figure~\ref{fig:resource-utilization} shows process memory and system free memory during a sustained-load test (200 requests at concurrency~10). The benchmark process RSS remains stable at $\approx$47~MB and heap usage at $\approx$7--8~MB, indicating no memory leaks. System free memory ($\approx$8.9~GB) is unaffected, confirming a minimal resource footprint.

Cold-start time---measured by launching MCP Bridge with all four servers from a stopped state---was 8,722~ms, dominated by Node.js module loading and MCP server initialization. Once started, subsequent tool calls incur only the REST latency described above. Additional benchmark figures (per-operation distributions, percentile heatmaps, CDF curves, API endpoint latency, and cold-start breakdown) are provided in Appendix~\ref{app:bench-supplementary}.

\FloatBarrier

\subsubsection{Bridge Overhead Analysis}
\begin{figure}[H]
\centering
\includegraphics[width=0.80\textwidth]{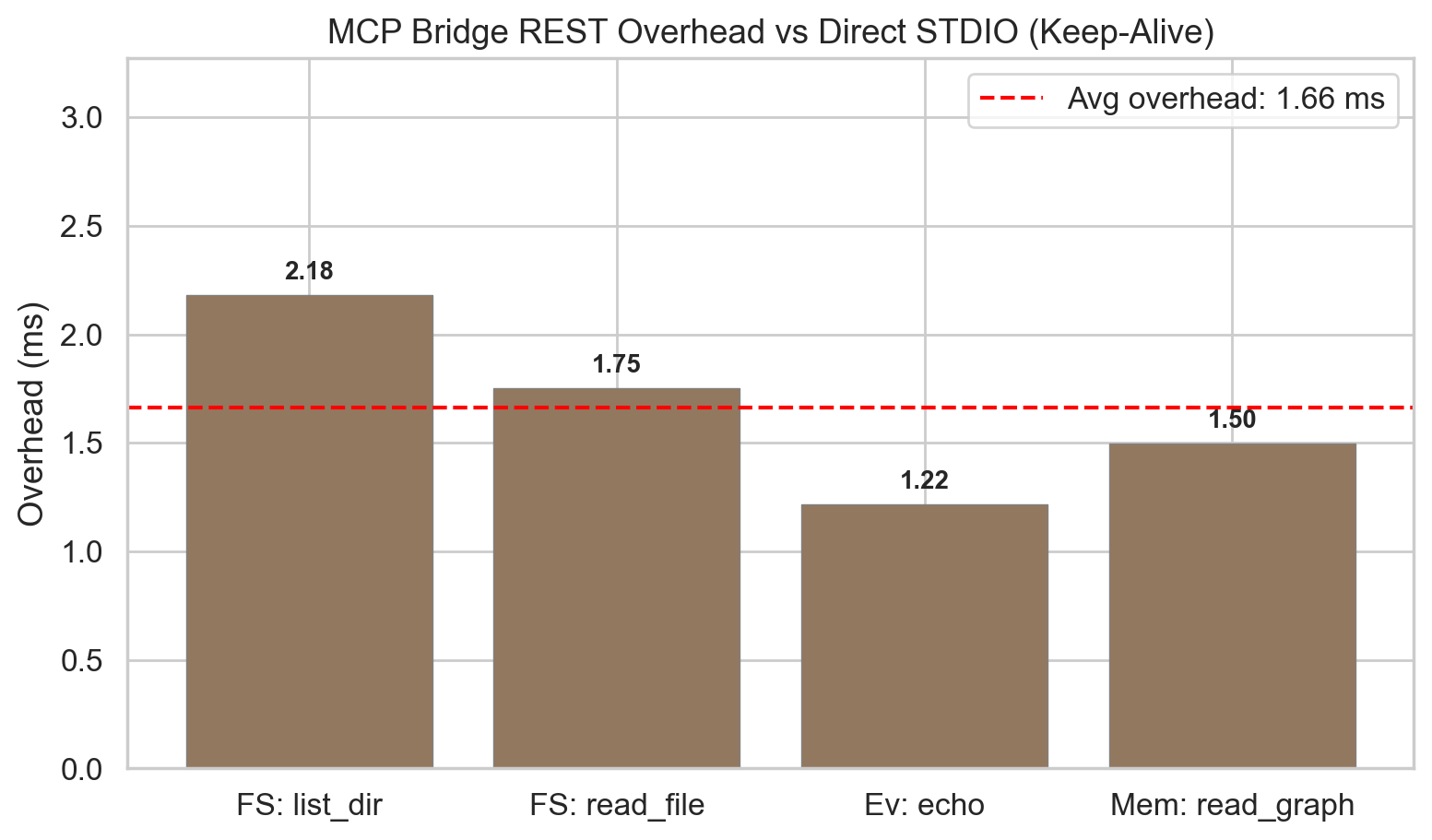}
\caption{Per-operation REST proxy overhead (Bridge REST minus STDIO keep-alive). The average overhead is 1.35~ms, attributable to HTTP parsing, JSON serialization, and Express.js routing.}\label{fig:bridge-overhead}
\end{figure}

Figure~\ref{fig:bridge-overhead} isolates the REST proxy overhead by subtracting STDIO keep-alive latency from Bridge REST latency for each operation. The mean overhead across operations is 1.35~ms (range: 1.07--1.64~ms), confirming that the HTTP/JSON serialization layer adds negligible cost relative to the convenience of network-accessible, multi-client MCP access.

\FloatBarrier

\section{Methodology: Policy Optimization for Tool Alignment}\label{sec4}

This section describes how we align open-weight Qwen3 models for reliable MCP-style tool use via reinforcement learning, while keeping the MCP Bridge interface model-agnostic. We focus on two practical objectives: selecting the correct tool(s) and emitting a parseable MCP-compliant tool-call structure that MCP Bridge can execute.

\subsection{Problem Setup and Output Interface}
Each example consists of a natural-language query $q$, a set of available tools $\mathcal{T}$ (tool name and description, optionally schema), and a reference set of expected tool names $\mathcal{T}^\star$. The model produces a response $y$ that may include one or more tool calls. In training and evaluation, tool calls are expected to appear as JSON enclosed in explicit tags:

\begin{verbatim}
<tool_call>
{"name": "tool_name", "arguments": {"arg1": "...", "arg2": "..."}}
</tool_call>
\end{verbatim}

We extract the predicted tool set $\widehat{\mathcal{T}}(y)$ from the model output and compare it to $\mathcal{T}^\star$. Tool calls are extracted using robust pattern matching over (i) tagged \texttt{<tool\_call>} blocks and (ii) common JSON-like formats produced by LLMs during tool use. To avoid false mismatches arising from server prefixes and naming conventions, tool names are normalized by lowercasing, removing known server prefixes (e.g., \texttt{Server::tool\_name} $\rightarrow$ \texttt{tool\_name}), stripping common MCP prefixes/suffixes, and
canonicalizing separators (hyphens/underscores/spaces) before matching.

\subsection{Training and Evaluation Data}
We train on tool-use data derived from Toucan-1.5M \cite{xu2025toucan}. The training script filters the SFT portion to subsets that contain tool calls (specifically \texttt{single-turn-original}, \texttt{single-turn-diversify}, and \texttt{multi-turn}) \cite{xu2025toucan}. Each training instance provides a user \texttt{question}, a tool list \texttt{tools} (with per-tool \texttt{name} and \texttt{description}), and a comma-separated \texttt{target\_tools} field indicating the expected tool(s). During prompt construction, the tool list is inserted into the system message under an \texttt{Available Tools} header, with the number of tools and description length capped to reduce context length pressure.

We evaluate on MCPToolBench++ \cite{fan2025mcptoolbench++}, a large-scale benchmark organized by tool category (\emph{Browser, File System, Search, Map, Pay, Finance}). Each evaluation sample provides a query, tool metadata,
and ground-truth function-call labels specifying the expected tool name(s). For evaluation, each category is scored independently, and we report Precision, Recall, F1, and Accuracy aggregated across samples.

\subsection{Metrics}
Let $P = \widehat{\mathcal{T}}(y)$ be the predicted tool set after normalization and $G=\mathcal{T}^\star$ be the
ground-truth tool set. We compute Precision $=|P\cap G|/|P|$, Recall $=|P\cap G|/|G|$, and
$F1 = 2\cdot \text{Prec}\cdot \text{Rec}/(\text{Prec}+\text{Rec})$ (with standard zero-handling). Accuracy is defined as exact set match (i.e., $P=G$) for single-turn MCPToolBench++ tasks.

\subsection{Reward Signal Used for RL}
Policy optimization uses a scalar reward that combines (i) whether the model chose the correct tool(s) and (ii) whether the tool call is formatted in a parseable MCP-compatible structure. Concretely, the total reward is:
\[
r(y; q,\mathcal{T}) \;=\; r_{\text{sel}}(y) \;+\; r_{\text{fmt}}(y).
\]
The tool-selection component $r_{\text{sel}}$ is derived from tool-set overlap (Precision/Recall/F1) between the extracted tool set and the reference tool set; the training script uses a piecewise mapping that strongly rewards near-perfect F1 and penalizes missing or completely incorrect tool calls. The format component $r_{\text{fmt}}$ rewards generating a \texttt{<tool\_call>} block containing valid JSON with the required \texttt{name} field, and assigns lower or negative scores when the output is malformed or not parseable.

The training code also includes additional reward variants (e.g., stricter JSON validity checks, schema compliance, parameter completeness, and length penalties), but our experiments found that the best-performing configuration for the models reported in this paper uses only the tool-selection reward and the format reward.

\subsection{Policy Optimization Methods}
We compare four policy-optimization variants implemented in the training pipeline: GRPO \cite{shao2024deepseekmath}, Dr.\ GRPO \cite{liu2503understanding}, DAPO \cite{yu2025dapo}, and BNPO \cite{xiao2025bnpo}. All methods operate on on-policy samples and update the model using group-based relative advantages (multiple generations per prompt), which reduces variance without requiring an explicit learned critic. GRPO \cite{shao2024deepseekmath} provides the baseline group-relative objective. Dr.\ GRPO \cite{liu2503understanding} modifies the objective to address training pathologies observed in R1-zero-like recipes. DAPO \cite{yu2025dapo} introduces an asymmetric clipping variant (with an additional high clipping parameter) designed to preserve learning signal as policies improve. BNPO \cite{xiao2025bnpo} introduces a bounded normalization strategy that stabilizes updates under shifting reward distributions.

\subsection{Implementation Details (Training Configuration)}\label{subsec:impl-details}
We implement training using Unsloth \cite{unsloth} for efficient Qwen3 loading and LoRA adaptation, and TRL-style
policy optimization with group sampling. All training runs and ablation experiments were carried out on a single NVIDIA RTX 6000 Pro GPU (96~GB VRAM). Table~\ref{tab:train-config} summarizes the configuration used in our
training script.

\begin{table}[h]
\centering
\caption{Training configuration used for Qwen3 tool-use policy optimization in our pipeline.}
\label{tab:train-config}
\begin{tabular}{p{0.34\textwidth} p{0.60\textwidth}}
\toprule
Setting & Value \\
\midrule
Backbone models & Qwen3-4B and Qwen3-8B (Unsloth checkpoints) \cite{unsloth} \\
Max sequence length & 2048 \\
Quantization & 4-bit loading enabled \\
LoRA rank ($r$) & 32 \\
LoRA target modules & \texttt{q\_proj, k\_proj, v\_proj, o\_proj, gate\_proj, up\_proj, down\_proj} \\
LoRA alpha & $2r$ (64) \\
Optimizer / scheduler & AdamW 8-bit, linear schedule with warmup ratio 0.1 \\
Learning rate / weight decay & $5\times 10^{-6}$ / 0.001 \\
Batching & per-device batch size 1, gradient accumulation 1 \\
On-policy generations & 4 samples per prompt (group size) \\
Sampling parameters & temperature 1.0; top-$p$ 1.0; min-$p$ 0.1; EOS stop enabled \\
Training steps / checkpointing & max steps 500; save every 200 steps \\
Reward functions used & tool-selection + format rewards (other reward variants tested but not used in final runs) \\
\bottomrule
\end{tabular}
\end{table}

\FloatBarrier

\section{Experiments and Results}\label{sec5}

We evaluate tool-use behavior on MCPToolBench++ \cite{fan2025mcptoolbench++}, reporting Precision, Recall, F1, and Accuracy for six tool categories (Browser, File System, Search, Map, Pay, Finance), with 50 randomly sampled evaluation instances per category ($N{=}300$ total; see Table~\ref{tab:eval-sizes}). We compare the base Qwen3 models against four RL methods (GRPO, Dr.\ GRPO, DAPO, BNPO), trained on tool-use data derived from Toucan-1.5M \cite{xu2025toucan} using the tool-selection and format rewards described in Section~\ref{sec4}. All reported 95\% confidence intervals (CIs) are computed via 10{,}000-iteration bootstrap resampling of per-sample scores.\footnote{All per-sample evaluation results were regenerated from scratch to compute bootstrap confidence intervals; the underlying models and test set are unchanged.}

\subsection{Per-category F1 trends}
\begin{figure}[H]
\centering
\includegraphics[width=0.98\textwidth]{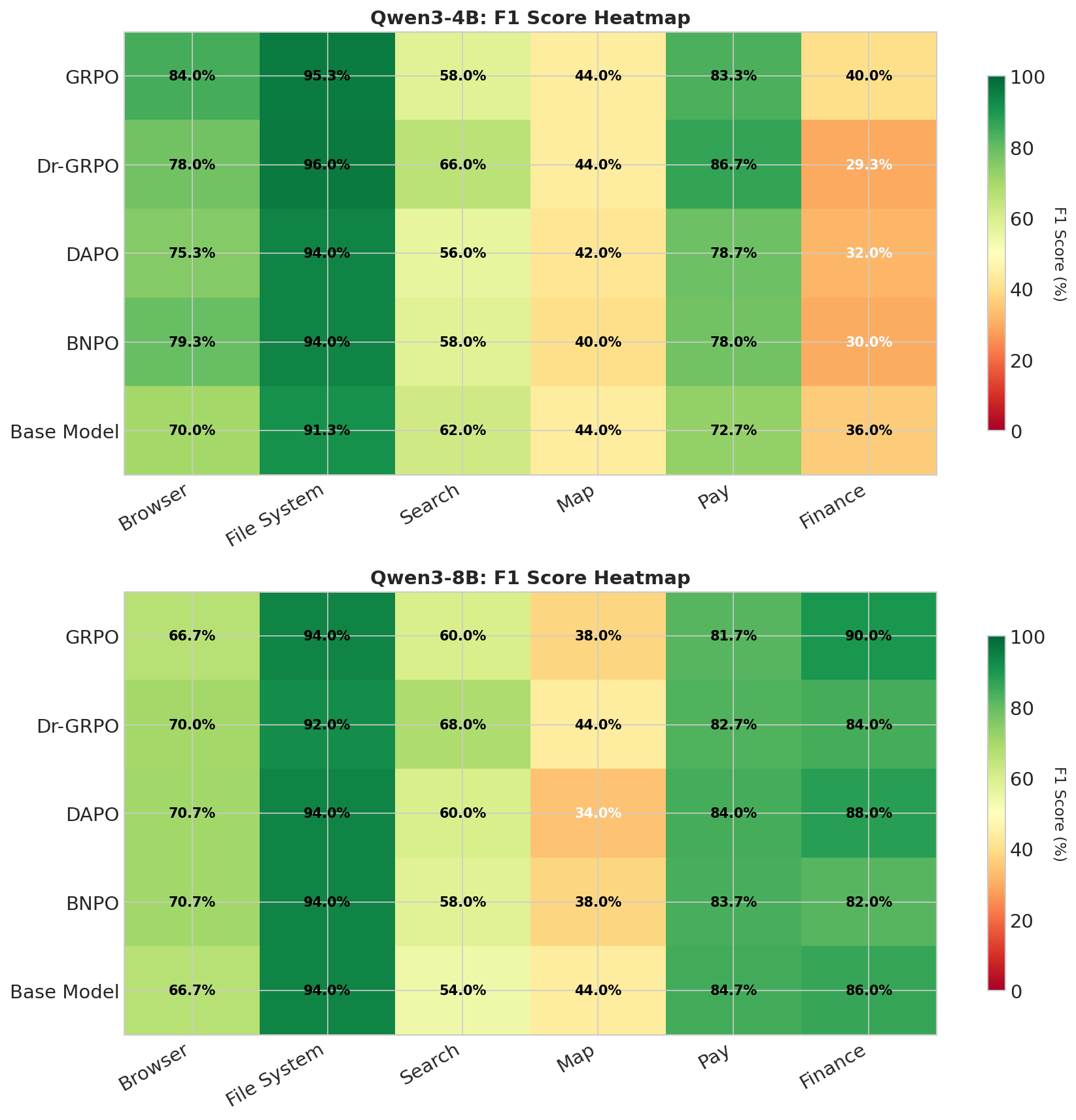}
\caption{Per-category F1 score heatmaps on MCPToolBench++ for Qwen3-4B (top) and Qwen3-8B (bottom). Rows correspond to RL methods and the base model; columns correspond to tool categories. Each cell reports mean F1 (\%) computed from normalized tool-name matching and robust tool-call extraction from model outputs. Darker green indicates better performance.}\label{fig:f1-heatmap}
\end{figure}

Figure~\ref{fig:f1-heatmap} summarizes per-category F1 scores. Two patterns appear consistently across model sizes. First, File System is the strongest category for all methods. Second, Map is persistently challenging across methods, with F1 clustered in a narrow band; this indicates that simply improving selection/formatting does not fully address category difficulty, likely due to higher semantic ambiguity in map-related intent and overlapping tool affordances.

\FloatBarrier

\begin{figure}[H]
\centering
\includegraphics[width=0.98\textwidth]{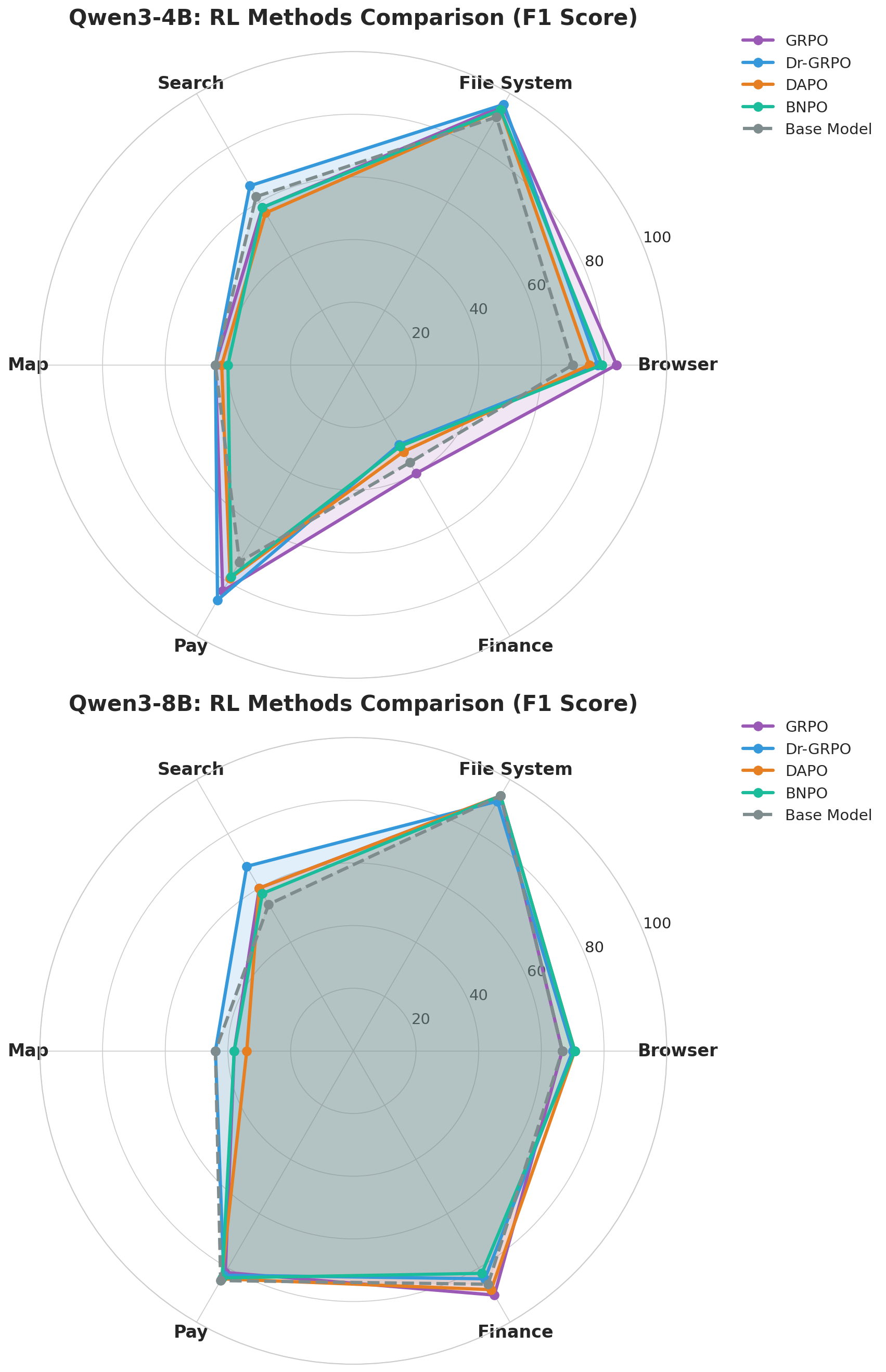}
\caption{Radar plots of per-category F1 on MCPToolBench++. This view emphasizes cross-category
trade-offs: improvements are not uniform across tools, and methods can help one category while degrading another (e.g.,
Finance sensitivity in smaller models).}\label{fig:f1-radar}
\end{figure}

A radar view of the same F1 values is shown in Figure~\ref{fig:f1-radar}. For Qwen3-4B, GRPO yields the largest gains, improving Browser from 70\% to 84\% and Pay from 72.7\% to 83.3\%; Dr.\ GRPO follows with the highest Search gain (66\% vs.\ 62\% base) and the best Pay (86.7\%). DAPO and BNPO show more moderate overall gains. For Qwen3-8B, Dr.\ GRPO produces the strongest result, raising Search from 54\% to 68\% and maintaining Finance (84\%); DAPO is close behind with the highest Browser (70.7\%), while GRPO achieves the best Finance (90\%) at a cost to Map and Pay.

\FloatBarrier

\subsection{Aggregate metrics and method comparison}
\begin{figure}[H]
\centering
\includegraphics[width=0.98\textwidth]{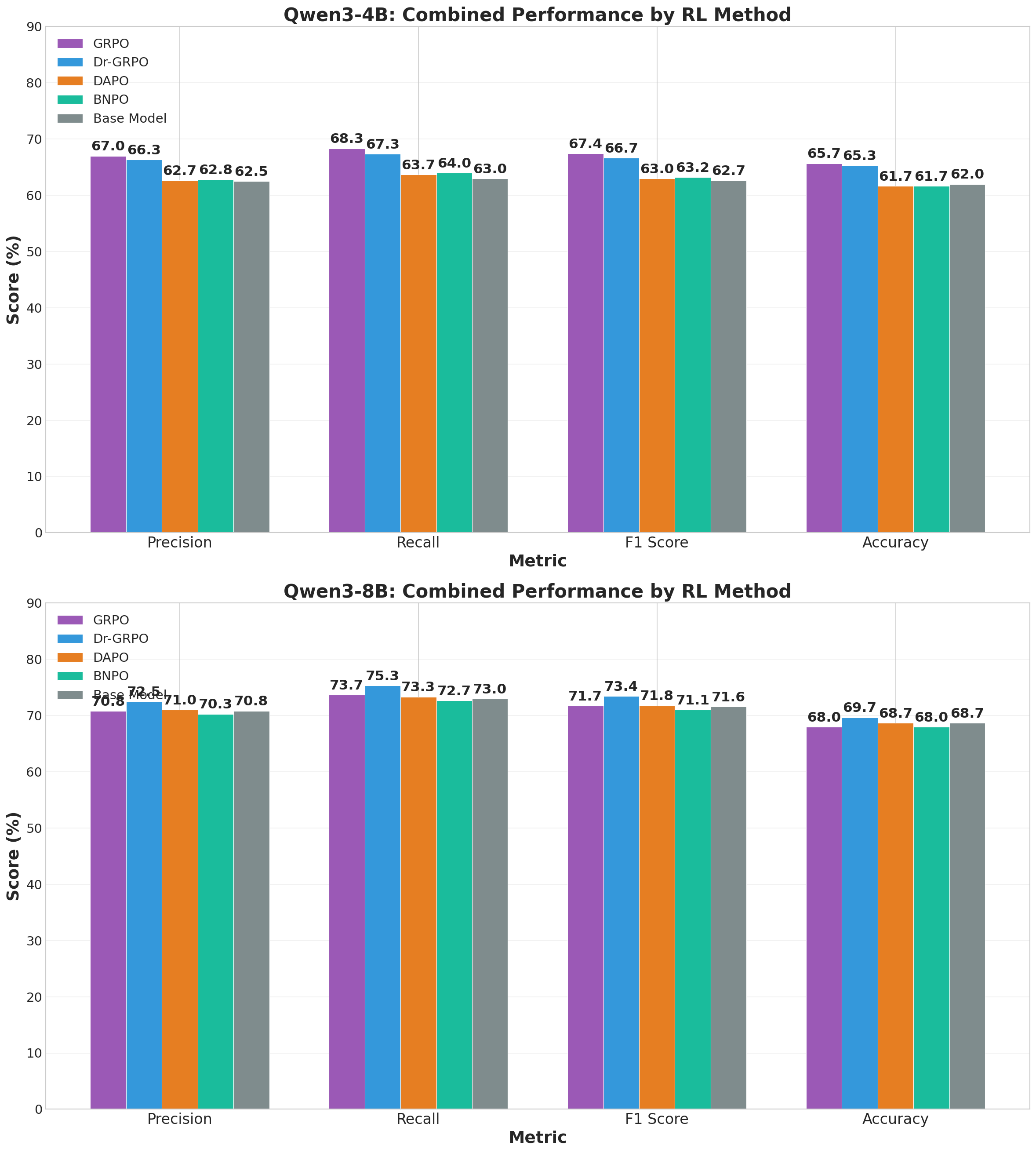}
\caption{Aggregate Precision, Recall, F1, and Accuracy (\%) on MCPToolBench++ \cite{fan2025mcptoolbench++} for Qwen3-4B
(top) and Qwen3-8B (bottom). Scores are averaged across the six tool categories. GRPO is the strongest overall method for Qwen3-4B, while Dr.\ GRPO is strongest for Qwen3-8B under the same reward design.}
\label{fig:combined-metrics}
\end{figure}

Figure~\ref{fig:combined-metrics} reports aggregate Precision, Recall, F1, and Accuracy (averaged over categories); full per-variant results with 95\% CIs are in Table~\ref{tab:full-results}.
For Qwen3-4B, GRPO yields the strongest aggregate F1 (67.4\%, 95\% CI [62.2, 72.7]) among models trained with both rewards, improving tool selection reliability relative to the base model (F1 62.7\%, CI [57.2, 68.0]). Dr.\ GRPO is close behind (F1 66.7\%), indicating that reference-distribution regularization
also delivers meaningful gains when the reward signal is dominated by selection correctness and format validity.

For Qwen3-8B, Dr.\ GRPO achieves the highest aggregate F1 (73.4\%, CI [68.6, 78.2]) and the best Accuracy (69.7\%), indicating that as capacity
increases, reference-aware updates can convert the same reward signal into more consistent improvements across categories.

A failure-type analysis of the base models (Table~\ref{tab:failure-analysis}) sheds light on what RL training addresses. For the 4B base model, the dominant failure mode is ``no tool call'' (21.7\% of samples), where the model does not produce any parseable tool invocation. At 8B, this drops to 12.3\%, but ``wrong tool'' errors increase to 12.7\% (vs.\ 9.0\% at 4B), suggesting that larger models are better at producing well-formed tool calls but still struggle with tool disambiguation. An ablation study isolating the tool-selection reward and the format reward is provided in Appendix~\ref{app:reward-ablation}.

\FloatBarrier

\subsection{Comparison against other models}
Table~\ref{tab:external-model-comparison} compares our best MCP-aligned checkpoints to several representative baselines
using the same metric suite. All API baselines were served through Groq's inference API \cite{groq2024}, which hosts open-weight models (Llama-3.3-70B, Llama-4-Maverick, Llama-4-Scout, GPT-OSS-120B) as well as the proprietary Kimi-K2-Instruct endpoint. The comparison contextualizes the tool-use reliability of MCP-aligned Qwen3 models against
larger models that may have broader general capabilities. While the largest baselines achieve higher absolute scores,
the RL-aligned Qwen3 models narrow the gap substantially relative to their base versions, demonstrating that targeted
policy optimization with selection+format rewards can produce practical tool-use behavior at smaller scales.

Notably, the 8B Dr.\ GRPO model achieves higher accuracy (69.7\%) than the 120B GPT-OSS-120B baseline (58.7\%), with non-overlapping bootstrap CIs, confirming that our 8-billion-parameter RL-aligned model can match or exceed much larger models on this benchmark. The 4B GRPO model (65.7\%) likewise outperforms GPT-OSS-120B. While the largest open-weight baselines (e.g.\ Llama-3.3-70B at 81.0\%) still lead, the margins to Kimi-K2, Llama-4-Maverick, and Llama-4-Scout are narrower, indicating that the 8B model is competitive with these mid-range baselines.

\begin{table}[h]
\centering
\caption{Tool-use evaluation comparison on MCPToolBench++ ($N{=}300$; 50 samples $\times$ 6 categories). Values are percentages; 95\% CIs from 10{,}000-iteration bootstrap resampling.}
\label{tab:external-model-comparison}
\begin{tabular}{lcccc}
\toprule
Model & Prec.\ (\%) & Rec.\ (\%) & F1 (\%) & Acc.\ (\%) \\
\midrule
MCP Bridge (4B + GRPO) & 67.0 & 68.3 & 67.4 [62.2, 72.7] & 65.7 [60.3, 71.0] \\
MCP Bridge (8B + Dr.\ GRPO) & 72.5 & 75.3 & 73.4 [68.6, 78.2] & 69.7 [64.3, 74.7] \\
\midrule
GPT-OSS-120B & 62.0 & 65.3 & 63.1 [57.8, 68.4] & 58.7 [53.0, 64.3] \\
Kimi-K2-Instruct & 77.3 & 80.7 & 78.4 [73.9, 82.7] & 74.0 [69.0, 78.7] \\
Llama-3.3-70B & 81.8 & 82.7 & 82.1 [77.7, 86.2] & 81.0 [76.7, 85.3] \\
Llama-4-Maverick & 77.5 & 80.0 & 78.3 [73.8, 82.8] & 75.3 [70.7, 80.3] \\
Llama-4-Scout & 75.0 & 76.3 & 75.3 [70.4, 80.1] & 74.3 [69.3, 79.3] \\
\bottomrule
\end{tabular}
\end{table}
\FloatBarrier

\begin{table}[h]
\centering
\caption{Failure-type breakdown for base (un-tuned) Qwen3 models ($N{=}300$). ``No tool call'': no parseable tool invocation; ``Wrong tool'': valid call to incorrect tool; ``Fmt.\ heuristic'': output rescued by fallback regex extraction.}
\label{tab:failure-analysis}
\begin{tabular}{lccccc}
\toprule
Model & Correct & No tool call & Wrong tool & Partial & Fmt.\ heuristic \\
\midrule
4B Base & 62.0\% & 21.7\% & 9.0\% & 1.0\% & 6.3\% \\
8B Base & 68.7\% & 12.3\% & 12.7\% & 4.3\% & 2.0\% \\
\bottomrule
\end{tabular}
\end{table}
\FloatBarrier

\begin{table}[h]
\centering
\caption{Evaluation samples per tool category in our MCPToolBench++ subset.}
\label{tab:eval-sizes}
\begin{tabular}{lc}
\toprule
Category & $N$ \\
\midrule
Browser & 50 \\
File System & 50 \\
Search & 50 \\
Map & 50 \\
Pay & 50 \\
Finance & 50 \\
\midrule
Total & 300 \\
\bottomrule
\end{tabular}
\end{table}
\FloatBarrier

\section{Future Work}\label{sec6}

MCP Bridge establishes a practical foundation for LLM-agnostic tool integration, and our RL-aligned Qwen3 clients demonstrate that open-weight models can reliably operate within this interface. Several directions remain to strengthen both the deployment layer and the model-alignment layer.

On the systems side, MCP Bridge can be extended with production-oriented features such as stronger multi-tenancy support, richer observability (structured tracing of tool calls and outcomes), and adaptive caching for idempotent requests. While the current implementation supports STDIO and SSE-backed MCP servers, broader transport coverage and improved server lifecycle management (including persistent server pools and backpressure-aware scheduling) would improve throughput under heavy load. Security can be expanded beyond the current risk levels by introducing fine-grained access policies (e.g., per-tool or per-argument allow/deny rules), integrating standard authentication/authorization mechanisms for enterprise deployments, and supporting more specialized sandbox profiles for high-risk tool execution.

On the modeling side, our current RL signal focuses on correct tool selection and MCP-compliant formatting. A natural next step is to incorporate execution-grounded feedback where rewards reflect tool-call success, argument validity, and downstream task completion rather than tool identity alone. This includes learning to fill parameters accurately, handling tool errors robustly, and improving multi-step tool planning over longer horizons. Additionally, aligning policies to the risk-based execution model itself (e.g., learning when a user confirmation is required or how to request missing permissions) could further improve safety and usability when operating against real MCP server fleets.

Finally, we plan to broaden evaluation beyond tool-name matching to include end-to-end success metrics, multi-turn tool-use trajectories, and real-world MCP environments. This will help quantify generalization to unseen servers, unseen tool schemas, and compositional tool sequences, which are critical for deploying MCP-based agents at scale.

\section{Conclusion}\label{sec7}

This paper introduced MCP Bridge, a lightweight, LLM-agnostic RESTful proxy that addresses key limitations of direct connections to Model Context Protocol (MCP) servers—especially the reliance on local STDIO execution that is impractical for browsers, mobile clients, and edge deployments. MCP Bridge exposes MCP server capabilities through a unified API, supports both STDIO and SSE server backends, and provides a risk-based execution model with standard execution, confirmation workflows, and Docker isolation for high-risk tools.

Beyond the proxy layer, we showed that reliable MCP usage depends on the client model's ability to produce tool calls that are both correct and protocol-compliant. To support open-weight deployment, we fine-tuned Qwen3-4B and Qwen3-8B using four RL methods (GRPO, Dr.\ GRPO, DAPO, BNPO) on Toucan-1.5M and evaluated them on MCPToolBench++ ($N{=}300$, 50 samples per category). Across model sizes, RL produces consistent directional improvements in tool-use reliability, with Qwen3-4B + GRPO achieving 67.4\% F1 (95\% CI [62.2, 72.7]) and 65.7\% Accuracy, and Qwen3-8B + Dr.\ GRPO achieving 73.4\% F1 (95\% CI [68.6, 78.2]) and 69.7\% Accuracy. Bootstrap CIs confirm that the 8B model outperforms the GPT-OSS-120B baseline (non-overlapping 95\% CIs), although base-to-RL improvements within either model size show overlapping intervals on the current evaluation set.

Taken together, MCP Bridge and RL-aligned open-weight clients provide a practical path toward deployable MCP-based agents: the proxy makes MCP server fleets accessible across platforms, and the aligned policies make open models reliable protocol clients. We hope this combination helps move MCP from a local-development interface toward a broadly deployable tool-access layer for heterogeneous AI systems.

\section*{Declarations}
\noindent\textbf{Funding:} This research received no specific grant from any funding agency in the public, commercial, or not-for-profit sectors.

\noindent\textbf{Clinical Trial Number:} Not applicable.

\noindent\textbf{Consent to Publish:} Not applicable.

\noindent\textbf{Data Availability:} The complete MCP Bridge implementation is available as an open-source project at \url{https://github.com/INQUIRELAB/mcp-bridge-api}.

\noindent\textbf{Ethics and Consent to Participate:} Not applicable.

\noindent\textbf{Competing Interests:} The authors declare that they have no competing interests.

\bibliography{sn-bibliography}

\newpage

\begin{appendices}

    \section{Reward Ablation Study}\label{app:reward-ablation}
    
    This appendix isolates the effect of the two reward components used in our RL training: the tool-selection reward and the format reward. We compare two ablations—\emph{Tool Select.\ Only} and \emph{Format Only}—against the main setting used in Section~\ref{sec5} (tool-selection + format). Table~\ref{tab:ablation-summary} summarizes the best aggregate results under each ablation.
    
    Across both Qwen3-4B and Qwen3-8B, the ablations reveal that the two reward components have complementary but asymmetric effects. Optimizing formatting alone can produce sizable gains: for 4B, the format-only ablation (BNPO, F1 67.1\%) is close to the full-reward setting (GRPO, F1 67.4\%), suggesting that at smaller scales, enforcing output structure alone captures most of the available improvement. For 8B, format-only (GRPO, F1 73.4\%) matches the full-reward result (Dr.\ GRPO, F1 73.4\%). In contrast, optimizing tool selection alone yields limited gains for both model sizes (4B: 61.3\%, 8B: 68.6\%), indicating that selection-only training is insufficient without explicit format incentives. The best overall performance is achieved when both rewards are used jointly, though the margin over format-only is modest.
    
    \begin{table}[h]
    \centering
    \caption{Best aggregate results (\%) under reward ablations. Each cell: F1 / Accuracy (best RL method in parentheses).}
    \label{tab:ablation-summary}
    \begin{tabular}{lccc}
    \toprule
    Model & Full (Sel.+Fmt.) & Tool Sel.\ Only & Format Only \\
    \midrule
    Qwen3-4B & 67.4 / 65.7 (GRPO) & 61.3 / 59.3 (DAPO) & 67.1 / 65.3 (BNPO) \\
    Qwen3-8B & 73.4 / 69.7 (Dr.\ GRPO) & 68.6 / 65.3 (Dr.\ GRPO) & 73.4 / 69.7 (GRPO) \\
    \bottomrule
    \end{tabular}
    \end{table}
    \FloatBarrier
    
    \begin{figure}[H]
    \centering
    \includegraphics[width=0.98\textwidth]{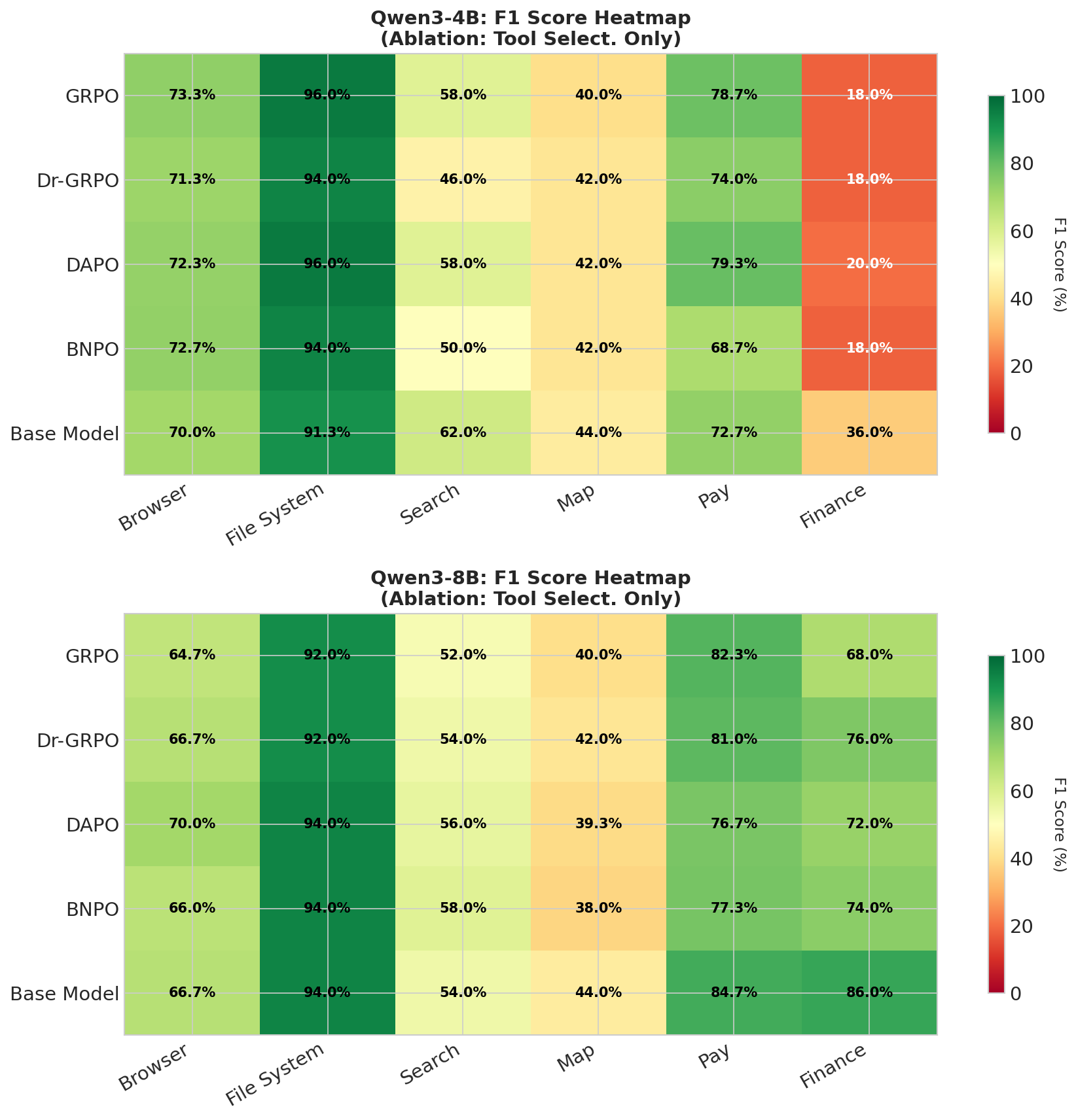}
    \caption{Per-category F1 heatmaps under the \emph{Tool Select. Only} ablation for Qwen3-4B (top) and Qwen3-8B (bottom). This ablation evaluates how much performance can be recovered when the reward focuses only on selecting the correct tool(s), without explicitly rewarding MCP-compliant formatting.}
    \label{fig:ablation-heatmap-select}
    \end{figure}
    \FloatBarrier
    
    \begin{figure}[H]
    \centering
    \includegraphics[width=0.98\textwidth]{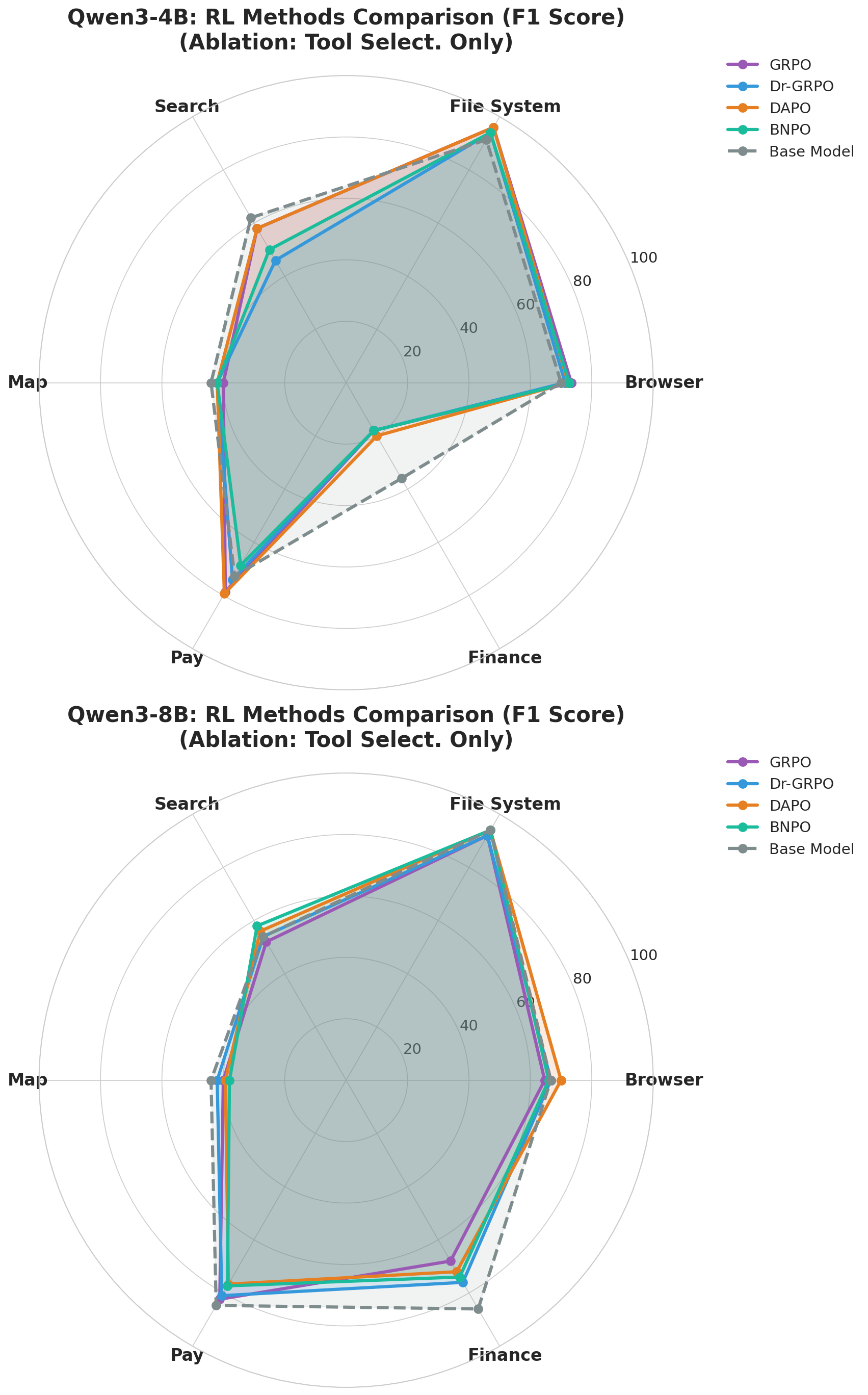}
    \caption{Radar plots of per-category F1 under the \emph{Tool Select. Only} ablation. The plots highlight that selection-only training does not uniformly improve categories and can degrade specific domains (notably Finance for smaller models), motivating the joint reward used in the main experiments.}
    \label{fig:ablation-radar-select}
    \end{figure}
    \FloatBarrier
    
    \begin{figure}[H]
    \centering
    \includegraphics[width=0.98\textwidth]{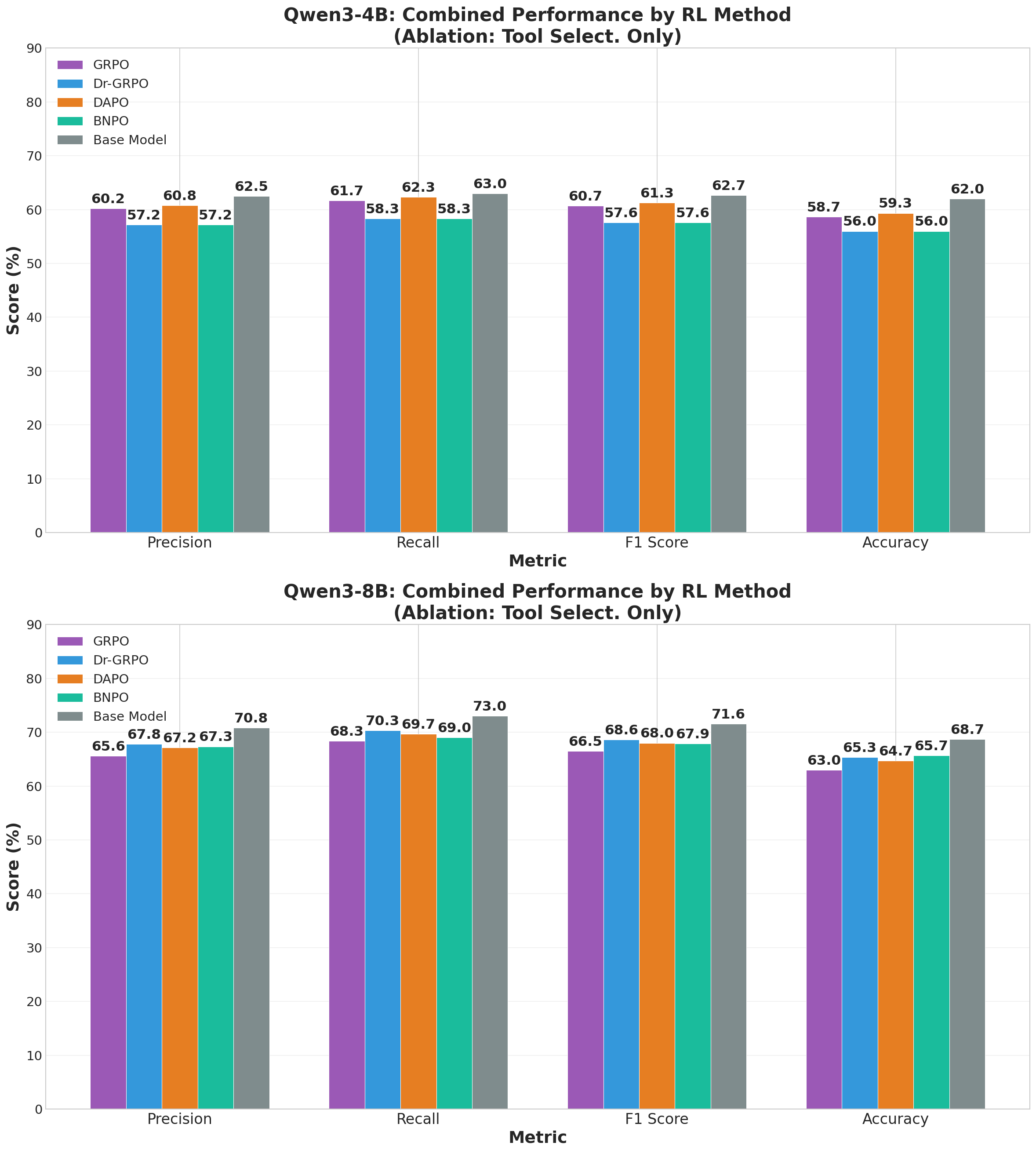}
    \caption{Aggregate Precision/Recall/F1/Accuracy under the \emph{Tool Select. Only} ablation. Compared to the main setting (Section~\ref{sec5}), selection-only training yields modest gains at 8B and limited or negative gains at 4B, indicating that explicit formatting incentives are important for stable improvements.}
    \label{fig:ablation-agg-select}
    \end{figure}
    \FloatBarrier
    
    \begin{figure}[H]
    \centering
    \includegraphics[width=0.98\textwidth]{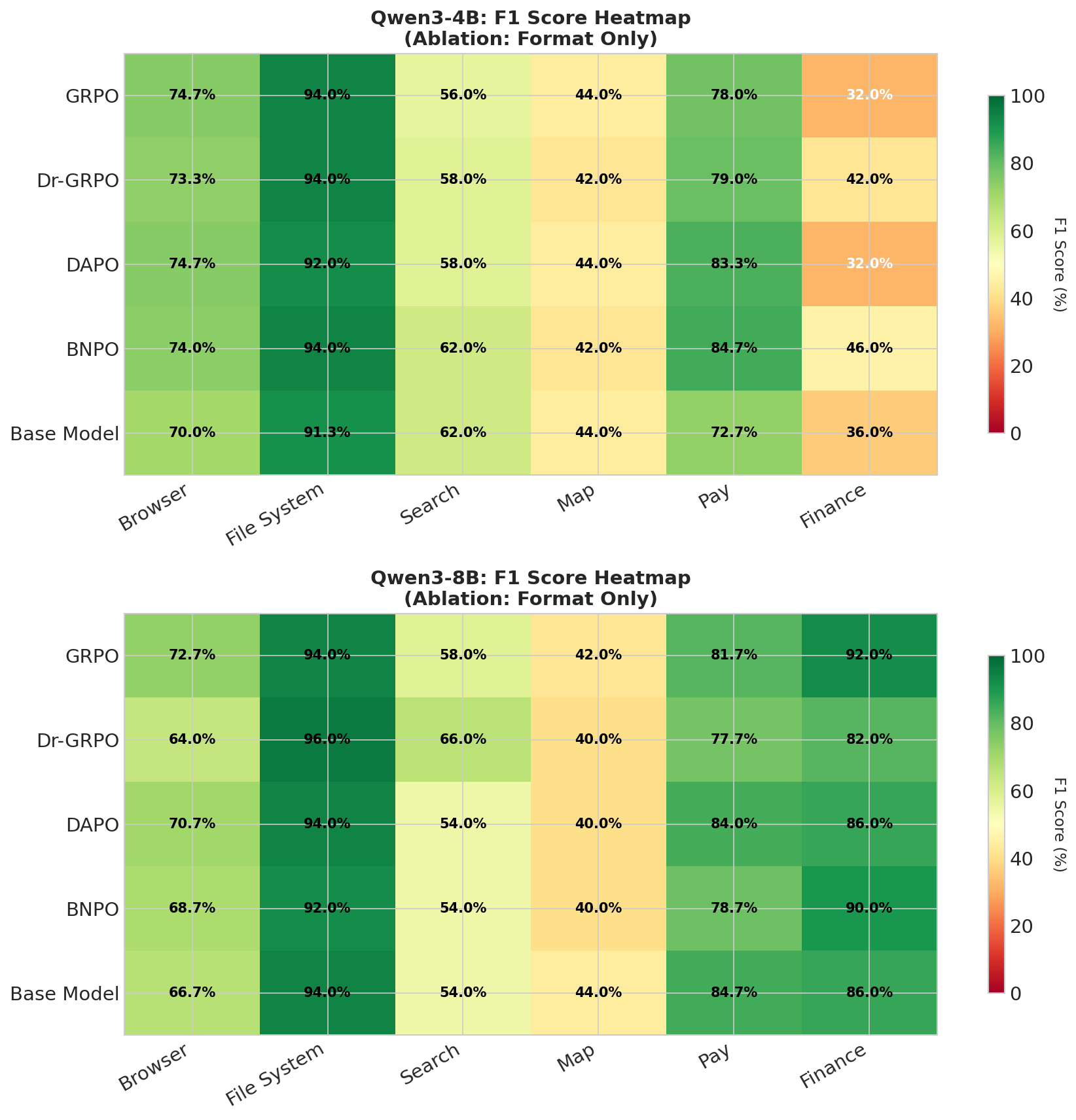}
    \caption{Per-category F1 heatmaps under the \emph{Format Only} ablation for Qwen3-4B (top) and Qwen3-8B (bottom). This ablation rewards emitting a parseable MCP tool-call structure without directly optimizing tool identity, testing how much of tool-use reliability comes from enforcing output structure alone.}
    \label{fig:ablation-heatmap-format}
    \end{figure}
    \FloatBarrier
    
    \begin{figure}[H]
    \centering
    \includegraphics[width=0.98\textwidth]{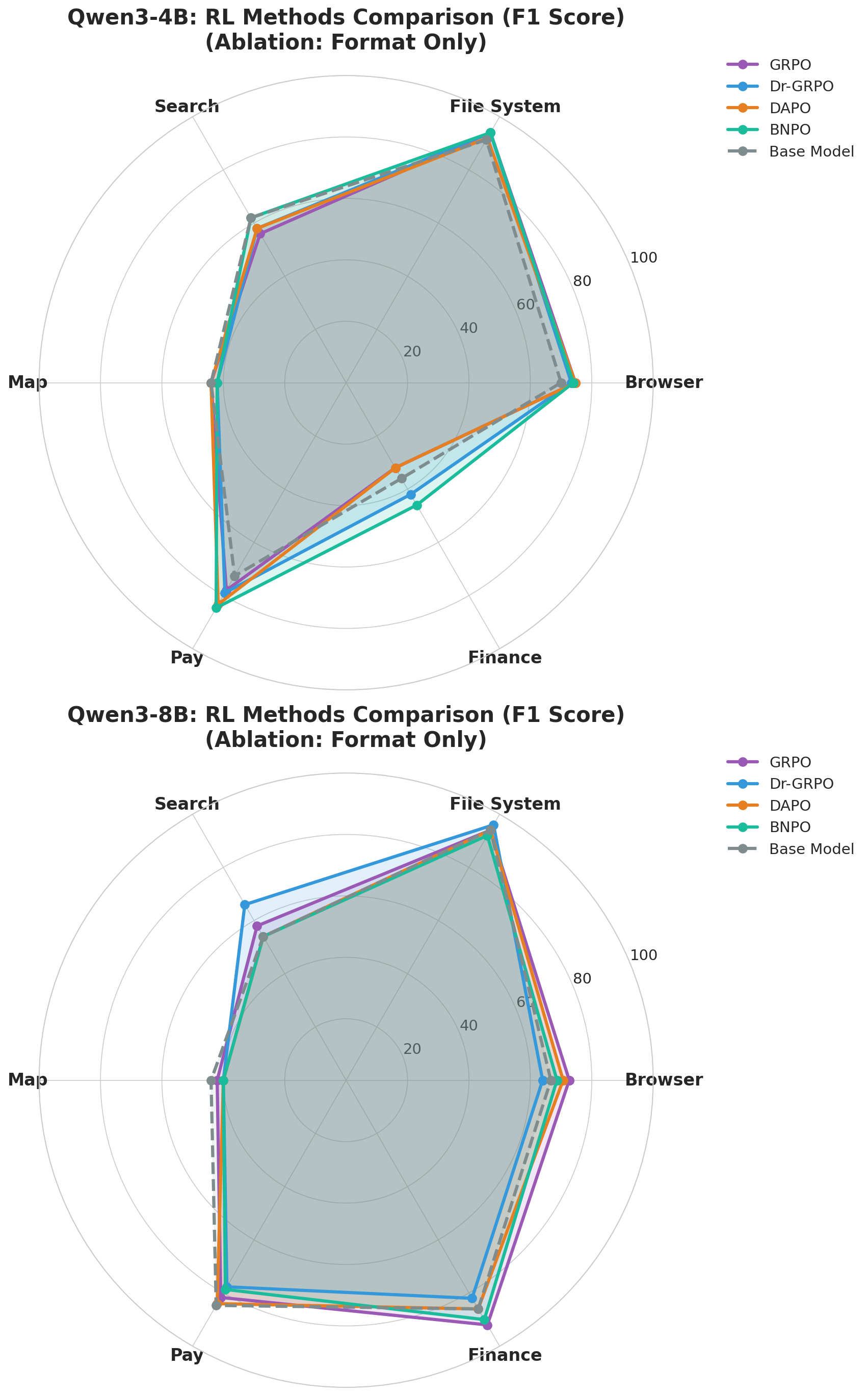}
    \caption{Radar plots of per-category F1 under the \emph{Format Only} ablation. Format-only optimization improves consistency in several categories, but the strongest results still come from jointly optimizing both selection and formatting (Section~\ref{sec5}).}
    \label{fig:ablation-radar-format}
    \end{figure}
    \FloatBarrier
    
    \begin{figure}[H]
    \centering
    \includegraphics[width=0.98\textwidth]{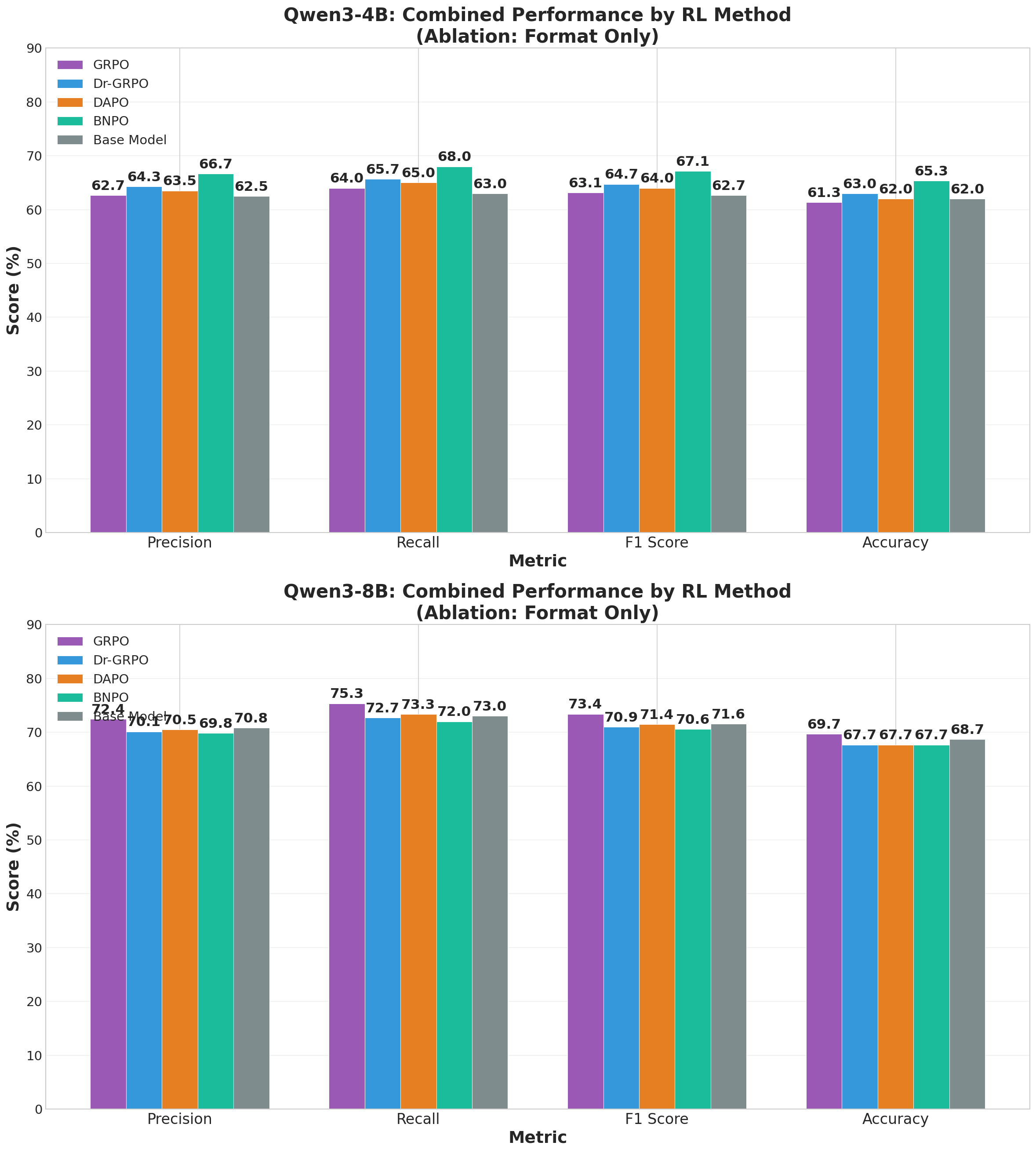}
    \caption{Aggregate Precision/Recall/F1/Accuracy under the \emph{Format Only} ablation. Format-only training can produce sizable gains (especially for Qwen3-4B), but combining format and tool-selection rewards yields the best overall metrics and the most reliable tool-use behavior.}
    \label{fig:ablation-agg-format}
    \end{figure}
    \FloatBarrier
    
    \section{Complete Evaluation Results}\label{app:full-results}
    
    Table~\ref{tab:full-results} reports the full evaluation results for all 26 local model variants and 5 API baselines on MCPToolBench++ ($N{=}300$). API baselines were served via Groq \cite{groq2024}. For local models, F1 and Accuracy are reported with 95\% bootstrap CIs (10{,}000 iterations). Rows are grouped by model size and reward configuration, sorted by descending F1 within each group.
    
    \begin{table}[h]
    \centering
    \caption{Complete evaluation results for all model variants on MCPToolBench++ ($N{=}300$).}
    \label{tab:full-results}
    \small
    \begin{tabular}{lcccc}
    \toprule
    Model & Prec.\ (\%) & Rec.\ (\%) & F1 (\%) [95\% CI] & Acc.\ (\%) [95\% CI] \\
    \midrule
    \multicolumn{5}{l}{\textit{Qwen3-4B}} \\
    \quad Base & 62.5 & 63.0 & 62.7 [57.2, 68.0] & 62.0 [56.3, 67.3] \\
    \quad GRPO (Full) & 67.0 & 68.3 & 67.4 [62.2, 72.7] & 65.7 [60.3, 71.0] \\
    \quad Dr.\ GRPO (Full) & 66.3 & 67.3 & 66.7 [61.4, 71.9] & 65.3 [60.0, 70.7] \\
    \quad BNPO (Full) & 62.8 & 64.0 & 63.2 [57.9, 68.6] & 61.7 [56.0, 67.0] \\
    \quad DAPO (Full) & 62.7 & 63.7 & 63.0 [57.6, 68.3] & 61.7 [56.0, 67.0] \\
    \quad DAPO (Sel.\ Only) & 60.8 & 62.3 & 61.3 [55.8, 66.7] & 59.3 [53.7, 64.7] \\
    \quad GRPO (Sel.\ Only) & 60.2 & 61.7 & 60.7 [55.2, 66.0] & 58.7 [53.0, 64.3] \\
    \quad BNPO (Sel.\ Only) & 57.2 & 58.3 & 57.6 [52.1, 63.0] & 56.0 [50.3, 61.7] \\
    \quad Dr.\ GRPO (Sel.\ Only) & 57.2 & 58.3 & 57.6 [52.0, 63.0] & 56.0 [50.3, 61.7] \\
    \quad BNPO (Fmt.\ Only) & 66.7 & 68.0 & 67.1 [61.9, 72.2] & 65.3 [60.0, 70.7] \\
    \quad Dr.\ GRPO (Fmt.\ Only) & 64.3 & 65.7 & 64.7 [59.4, 70.0] & 63.0 [57.7, 68.3] \\
    \quad DAPO (Fmt.\ Only) & 63.5 & 65.0 & 64.0 [58.6, 69.2] & 62.0 [56.7, 67.3] \\
    \quad GRPO (Fmt.\ Only) & 62.7 & 64.0 & 63.1 [57.8, 68.3] & 61.3 [55.7, 66.7] \\
    \midrule
    \multicolumn{5}{l}{\textit{Qwen3-8B}} \\
    \quad Base & 70.8 & 73.0 & 71.6 [66.6, 76.4] & 68.7 [63.3, 74.0] \\
    \quad Dr.\ GRPO (Full) & 72.5 & 75.3 & 73.4 [68.6, 78.2] & 69.7 [64.3, 74.7] \\
    \quad DAPO (Full) & 71.0 & 73.3 & 71.8 [66.6, 76.7] & 68.7 [63.3, 74.0] \\
    \quad GRPO (Full) & 70.8 & 73.7 & 71.7 [66.7, 76.6] & 68.0 [62.7, 73.3] \\
    \quad BNPO (Full) & 70.3 & 72.7 & 71.1 [66.0, 75.9] & 68.0 [62.7, 73.0] \\
    \quad Dr.\ GRPO (Sel.\ Only) & 67.8 & 70.3 & 68.6 [63.4, 73.8] & 65.3 [60.0, 70.7] \\
    \quad DAPO (Sel.\ Only) & 67.2 & 69.7 & 68.0 [62.8, 73.1] & 64.7 [59.3, 70.0] \\
    \quad BNPO (Sel.\ Only) & 67.3 & 69.0 & 67.9 [62.7, 73.0] & 65.7 [60.3, 71.0] \\
    \quad GRPO (Sel.\ Only) & 65.6 & 68.3 & 66.5 [61.2, 71.7] & 63.0 [57.7, 68.3] \\
    \quad GRPO (Fmt.\ Only) & 72.4 & 75.3 & 73.4 [68.4, 78.1] & 69.7 [64.3, 74.7] \\
    \quad DAPO (Fmt.\ Only) & 70.5 & 73.3 & 71.4 [66.4, 76.3] & 67.7 [62.3, 73.0] \\
    \quad Dr.\ GRPO (Fmt.\ Only) & 70.1 & 72.7 & 70.9 [65.9, 75.9] & 67.7 [62.3, 73.0] \\
    \quad BNPO (Fmt.\ Only) & 69.8 & 72.0 & 70.6 [65.4, 75.6] & 67.7 [62.3, 73.0] \\
    \midrule
    \multicolumn{5}{l}{\textit{API Baselines}} \\
    \quad GPT-OSS-120B & 62.0 & 65.3 & 63.1 [57.8, 68.4] & 58.7 [53.0, 64.3] \\
    \quad Kimi-K2-Instruct & 77.3 & 80.7 & 78.4 [73.9, 82.7] & 74.0 [69.0, 78.7] \\
    \quad Llama-3.3-70B & 81.8 & 82.7 & 82.1 [77.7, 86.2] & 81.0 [76.7, 85.3] \\
    \quad Llama-4-Maverick & 77.5 & 80.0 & 78.3 [73.8, 82.8] & 75.3 [70.7, 80.3] \\
    \quad Llama-4-Scout & 75.0 & 76.3 & 75.3 [70.4, 80.1] & 74.3 [69.3, 79.3] \\
    \bottomrule
    \end{tabular}
    \end{table}
    \FloatBarrier
    
    \section{MCP Bridge Benchmark --- Supplementary Figures}\label{app:bench-supplementary}
    
    This appendix provides additional benchmark visualizations that complement the system evaluation in Section~\ref{subsec3.6}. All measurements were collected under the same configuration described in Table~\ref{tab:bench-env} (4 MCP servers, 50 tools, Node.js v22.12.0, AMD Ryzen~7 7435HS, 24~GB RAM).
    
    \begin{figure}[H]
    \centering
    \includegraphics[width=0.90\textwidth]{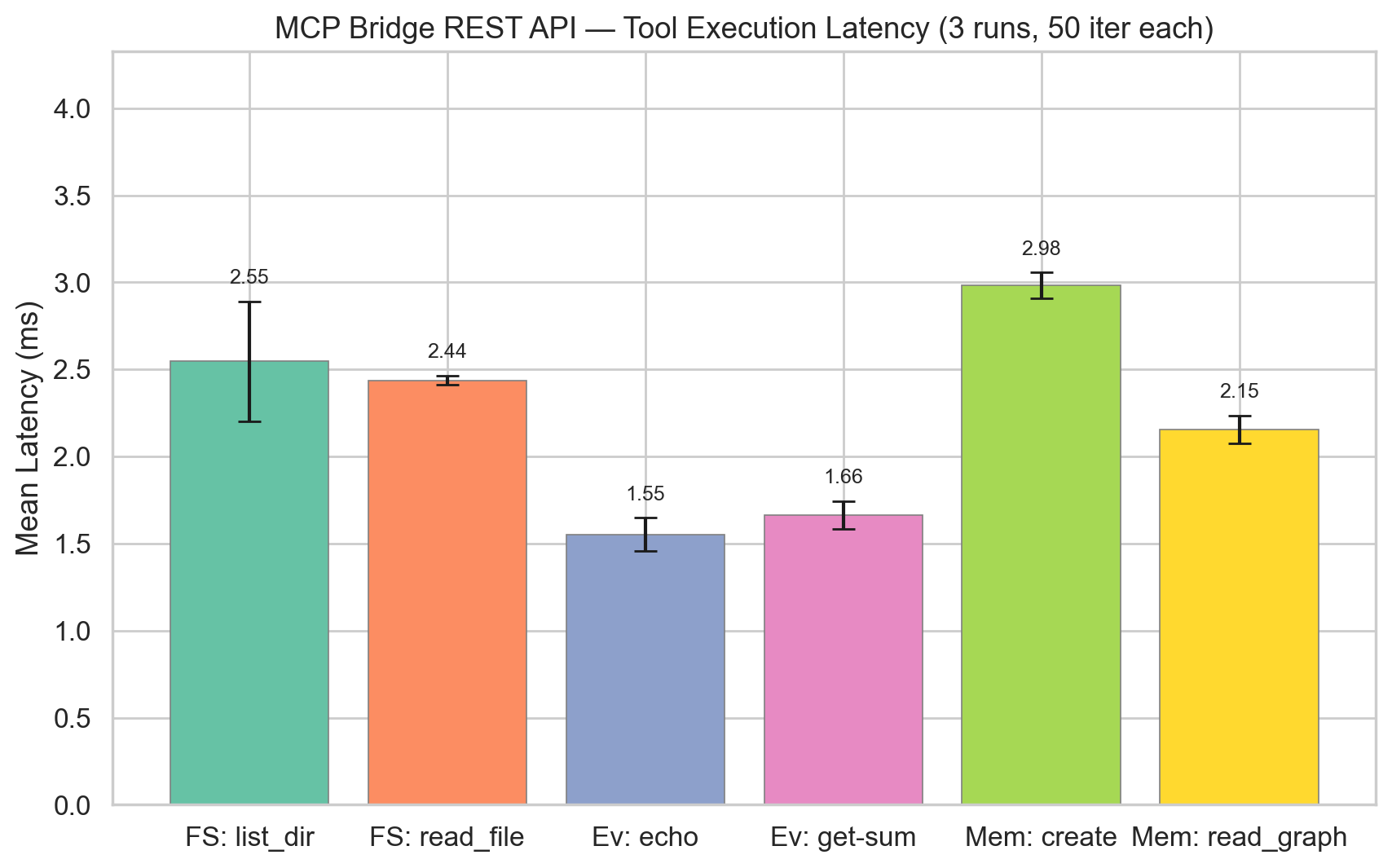}
    \caption{Bridge REST mean latency per tool-execution operation with error bars (mean~$\pm$~std, 3 runs, 50 iterations each). Lightweight operations such as \texttt{echo} and \texttt{get-sum} complete in under 2~ms.}
    \label{fig:supp-mean-latency}
    \end{figure}
    \FloatBarrier
    
    \begin{figure}[H]
    \centering
    \includegraphics[width=0.90\textwidth]{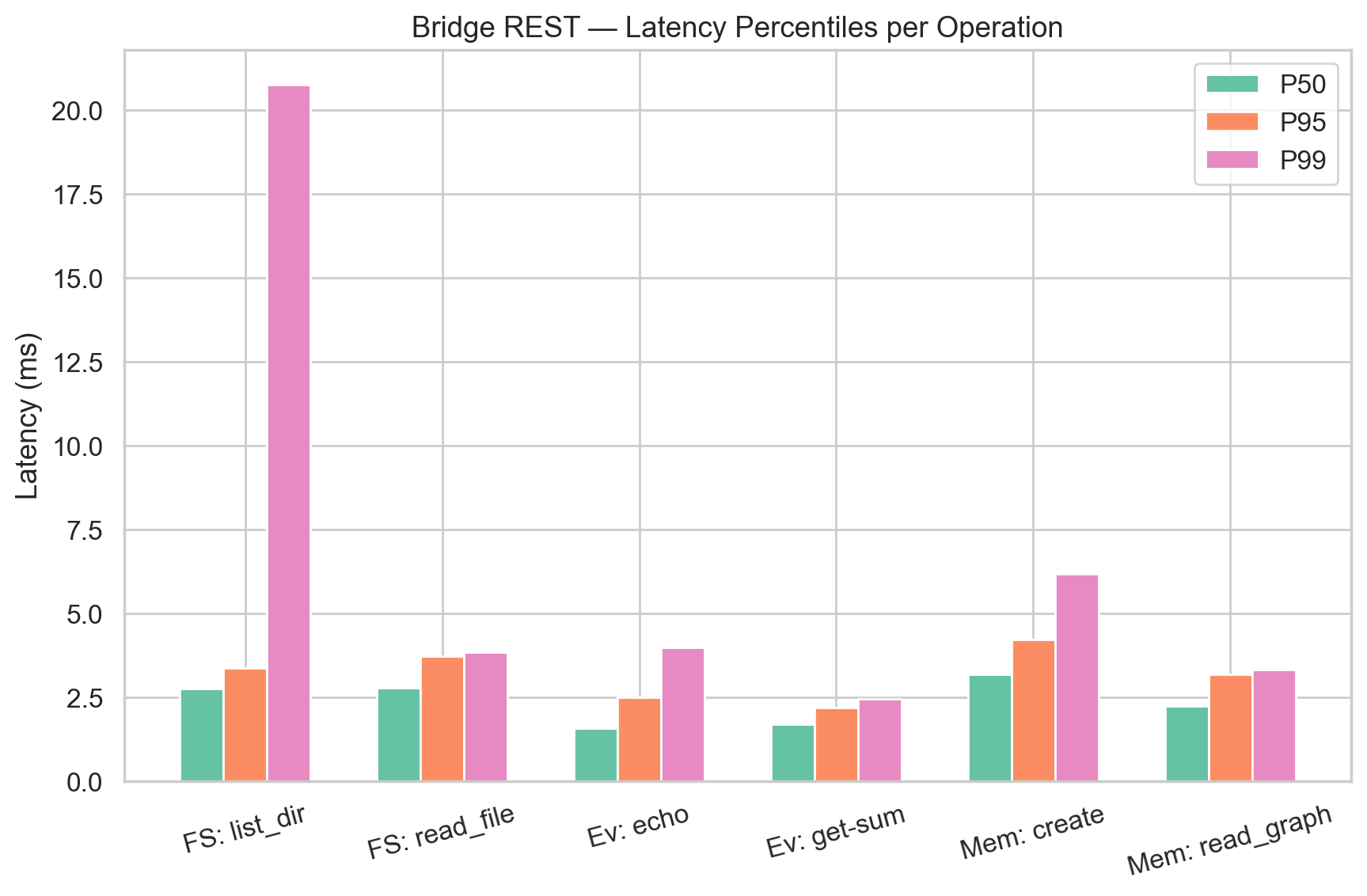}
    \caption{Latency percentile breakdown (P50, P95, P99) per operation. The gap between P50 and P99 is small for most operations, indicating low tail-latency variance.}
    \label{fig:supp-percentiles}
    \end{figure}
    \FloatBarrier
    
    \begin{figure}[H]
    \centering
    \includegraphics[width=0.90\textwidth]{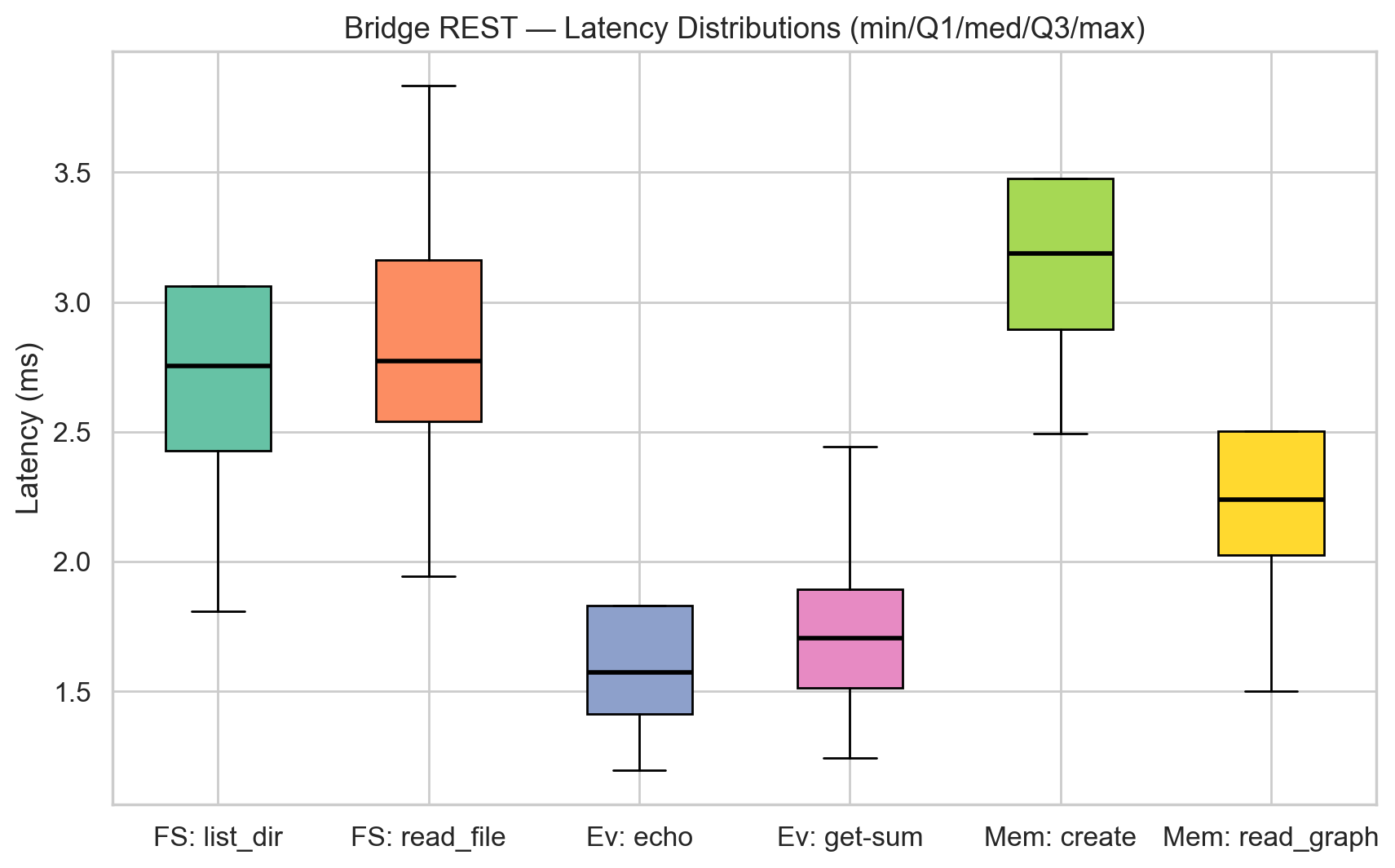}
    \caption{Latency distributions (min, Q1, median, Q3, max) per operation. The tight interquartile ranges confirm consistent response times across iterations.}
    \label{fig:supp-boxplot}
    \end{figure}
    \FloatBarrier
    
    \begin{figure}[H]
    \centering
    \includegraphics[width=0.90\textwidth]{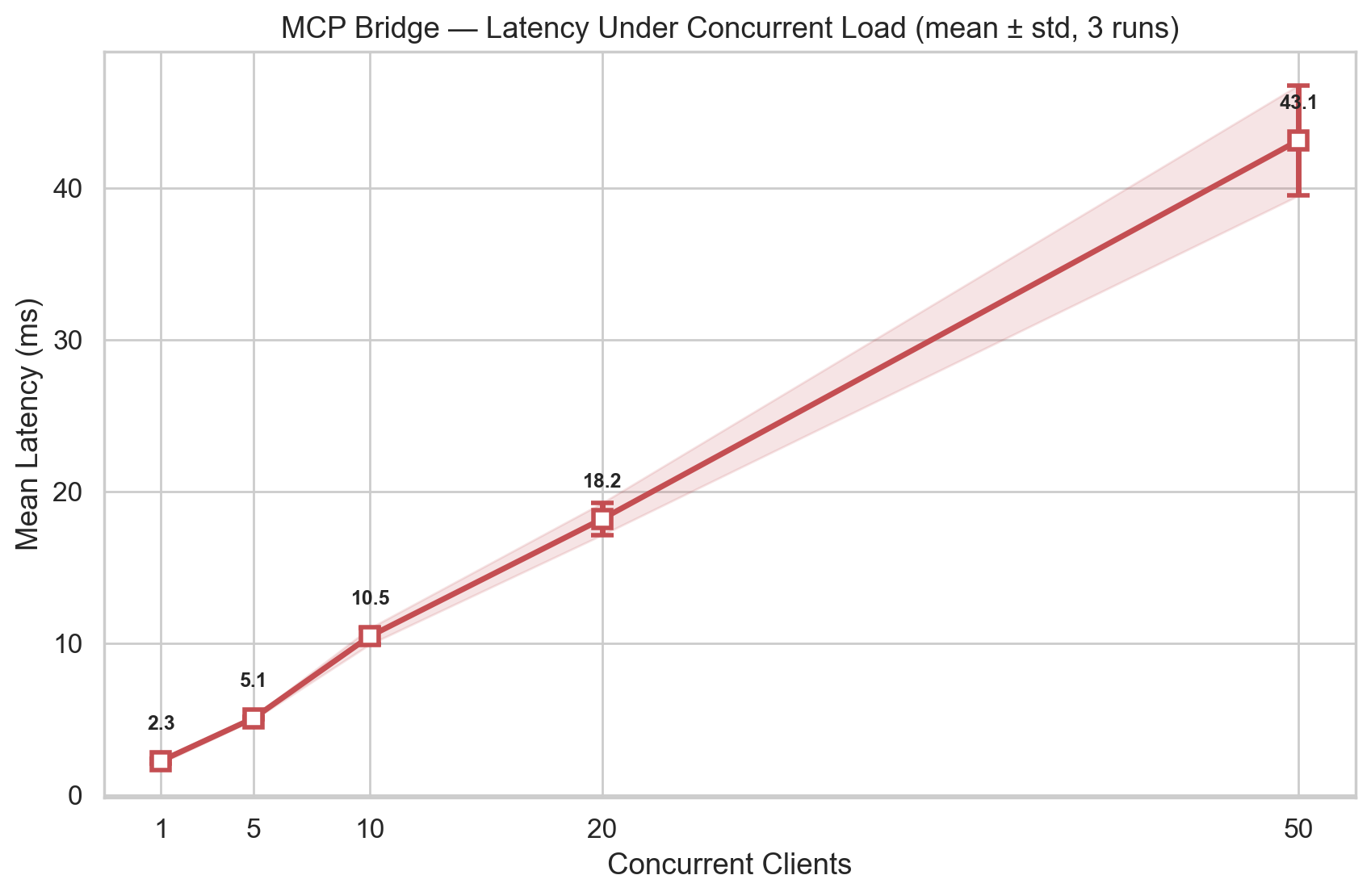}
    \caption{Mean latency as a function of concurrent clients (mean~$\pm$~std, 3 runs). Latency increases approximately linearly with concurrency, consistent with single-threaded event-loop behavior.}
    \label{fig:supp-latency-concurrency}
    \end{figure}
    \FloatBarrier
    
    \begin{figure}[H]
    \centering
    \includegraphics[width=0.90\textwidth]{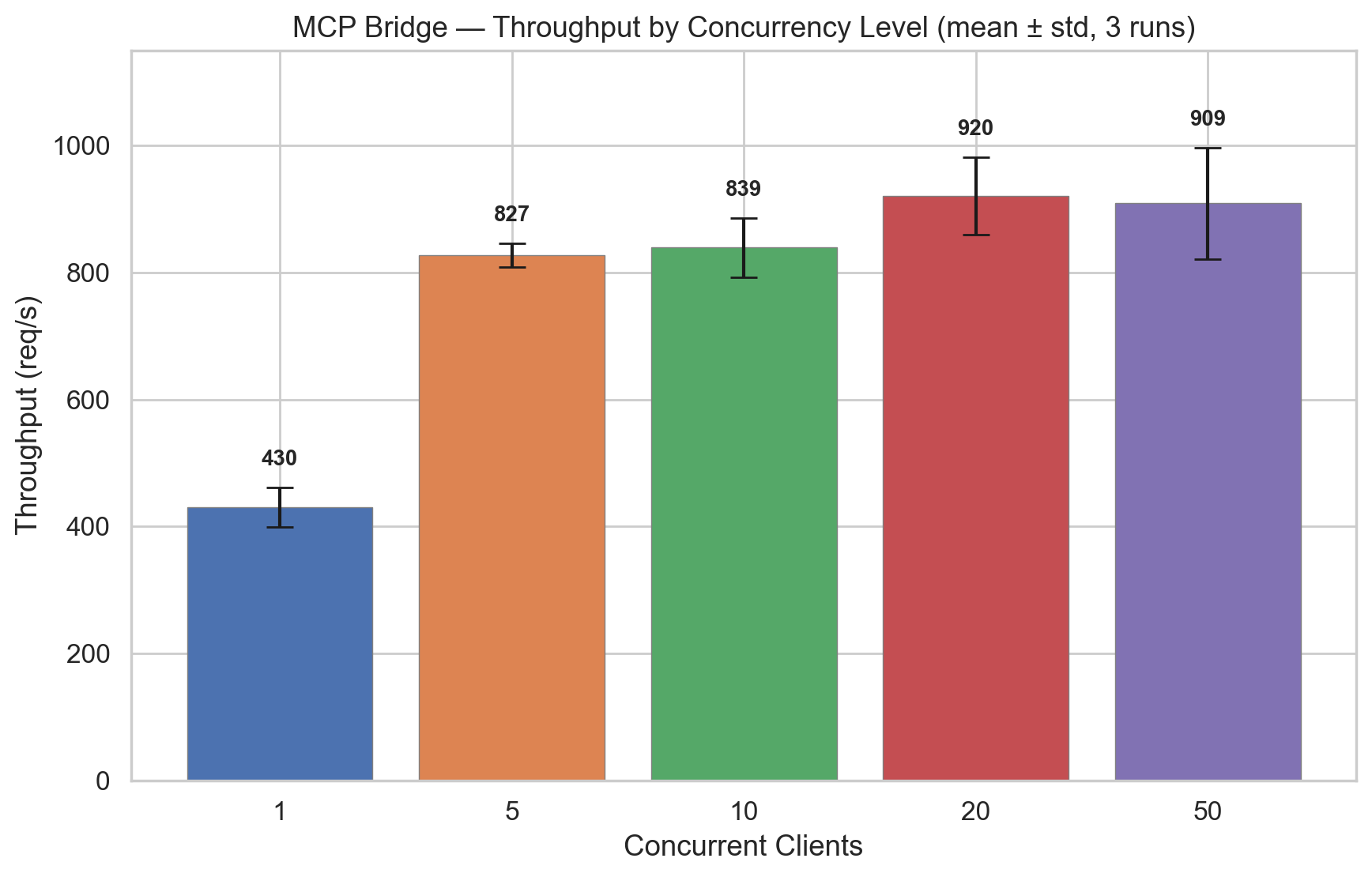}
    \caption{Throughput by concurrency level (bar chart with error bars, 3 runs). Throughput plateaus near 900~req/s at concurrency 20--50.}
    \label{fig:supp-throughput-bar}
    \end{figure}
    \FloatBarrier
    
    \begin{figure}[H]
    \centering
    \includegraphics[width=0.75\textwidth]{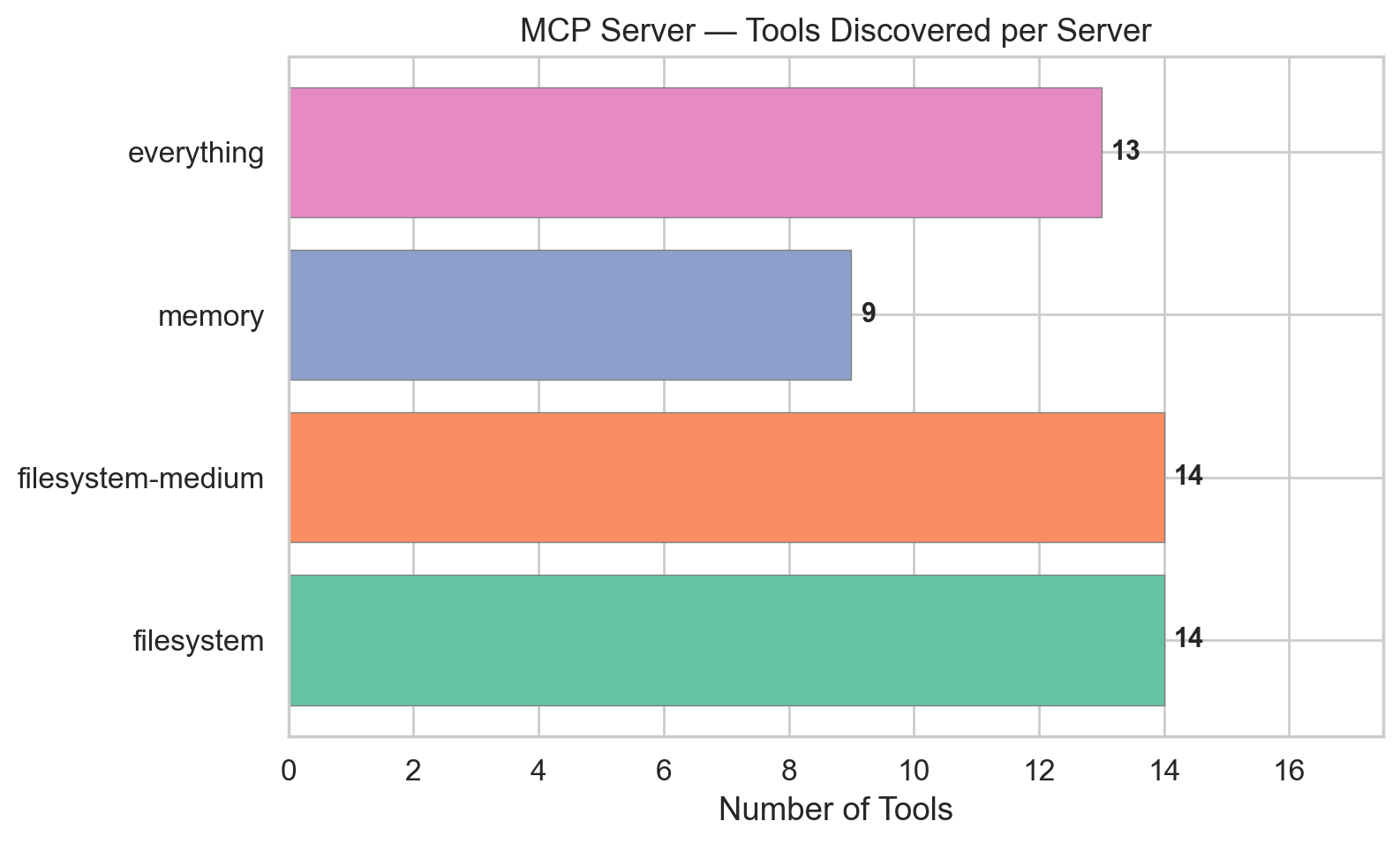}
    \caption{Number of tools discovered per MCP server. MCP Bridge's automatic tool discovery successfully enumerated all 50 tools across 4 servers upon initialization.}
    \label{fig:supp-tool-count}
    \end{figure}
    \FloatBarrier
    
    \begin{figure}[H]
    \centering
    \includegraphics[width=0.90\textwidth]{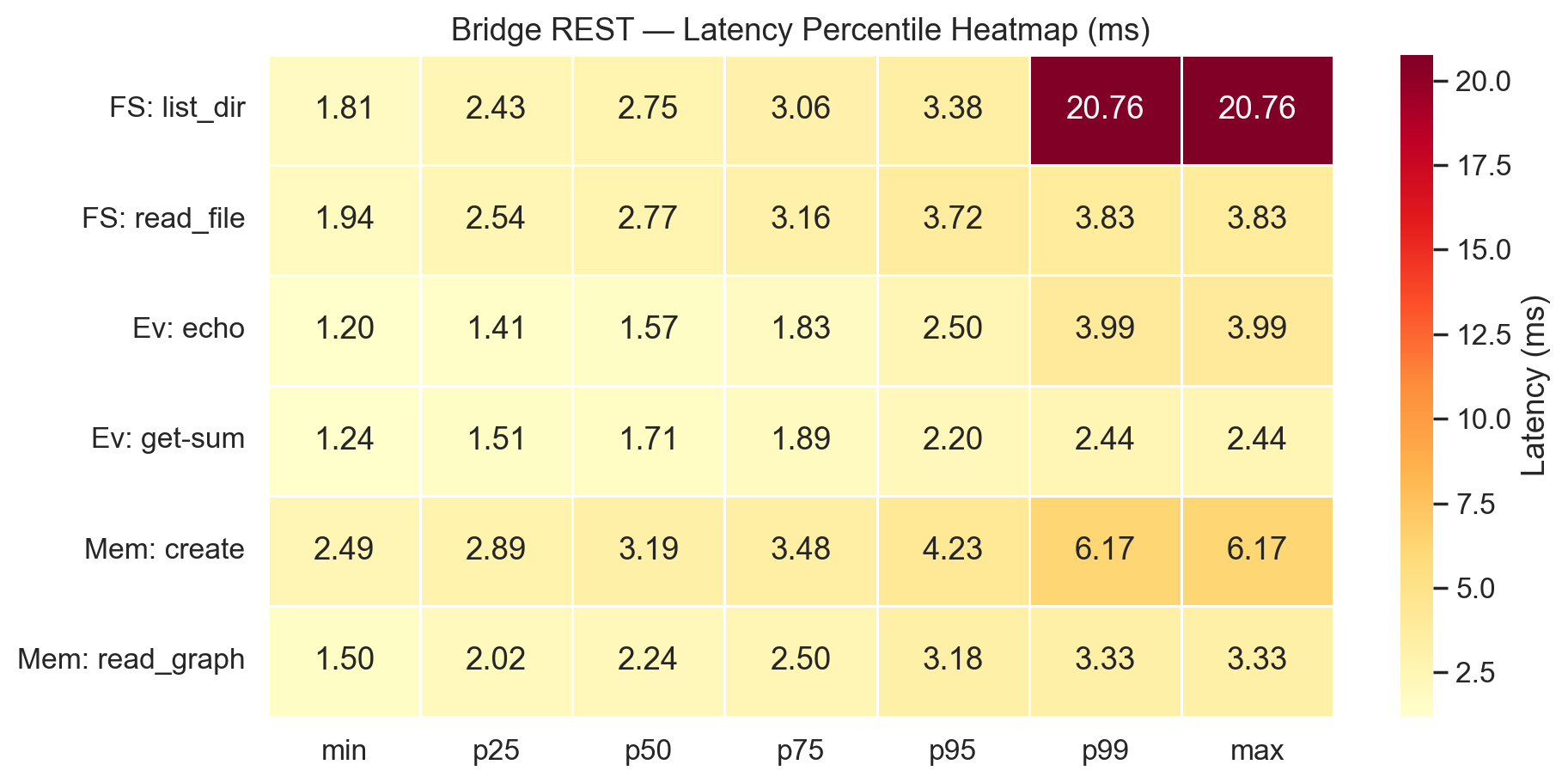}
    \caption{Latency percentile heatmap across all tool-execution operations. Cells report latency in milliseconds; darker shading indicates higher values.}
    \label{fig:supp-heatmap}
    \end{figure}
    \FloatBarrier
    
    \begin{figure}[H]
    \centering
    \includegraphics[width=0.90\textwidth]{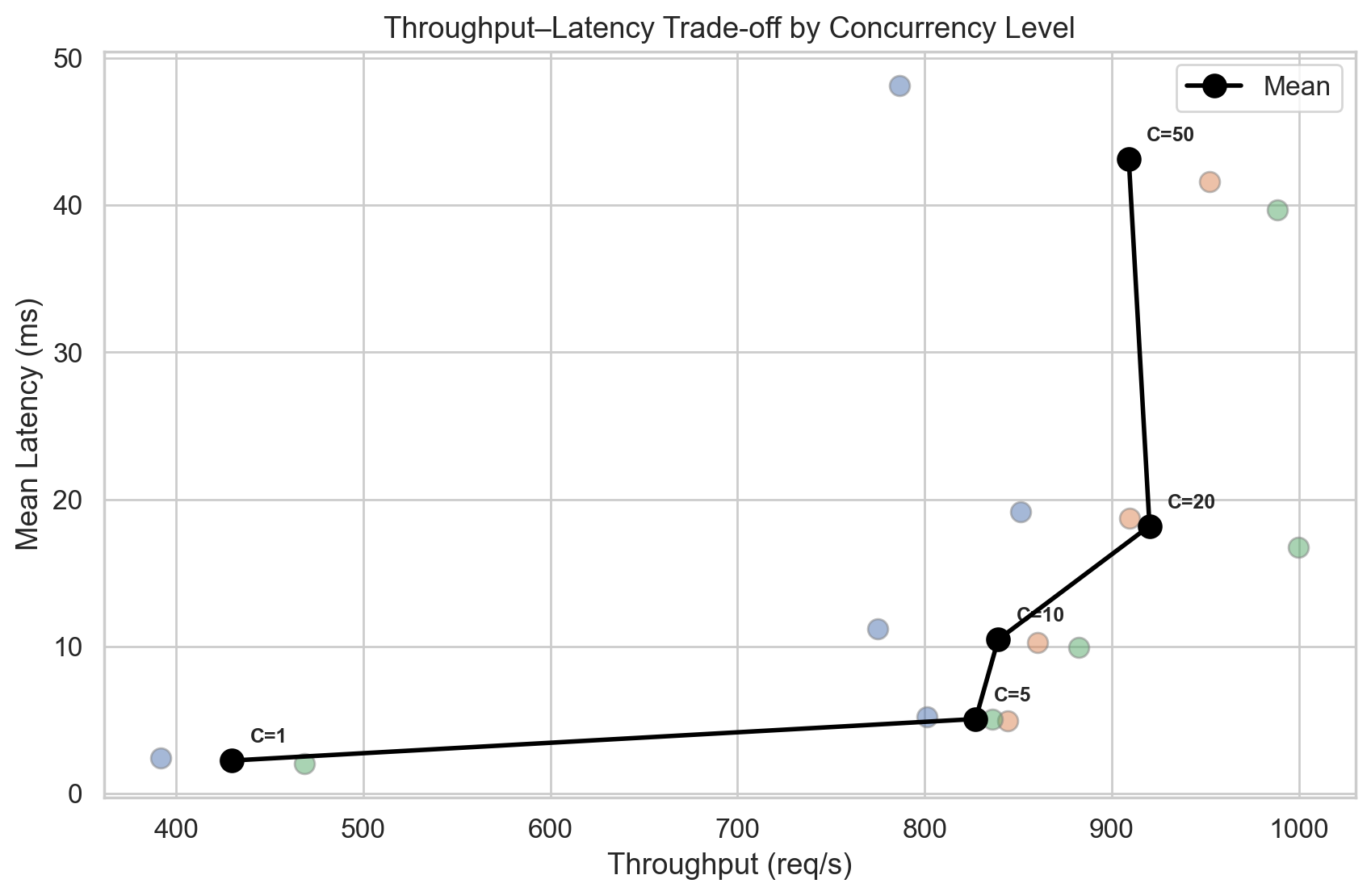}
    \caption{Throughput--latency trade-off by concurrency level. Individual data points from 3 runs are shown alongside the connected mean trajectory, illustrating the trade-off between higher throughput and increased per-request latency.}
    \label{fig:supp-tp-lat-scatter}
    \end{figure}
    \FloatBarrier
    
    \begin{figure}[H]
    \centering
    \includegraphics[width=0.75\textwidth]{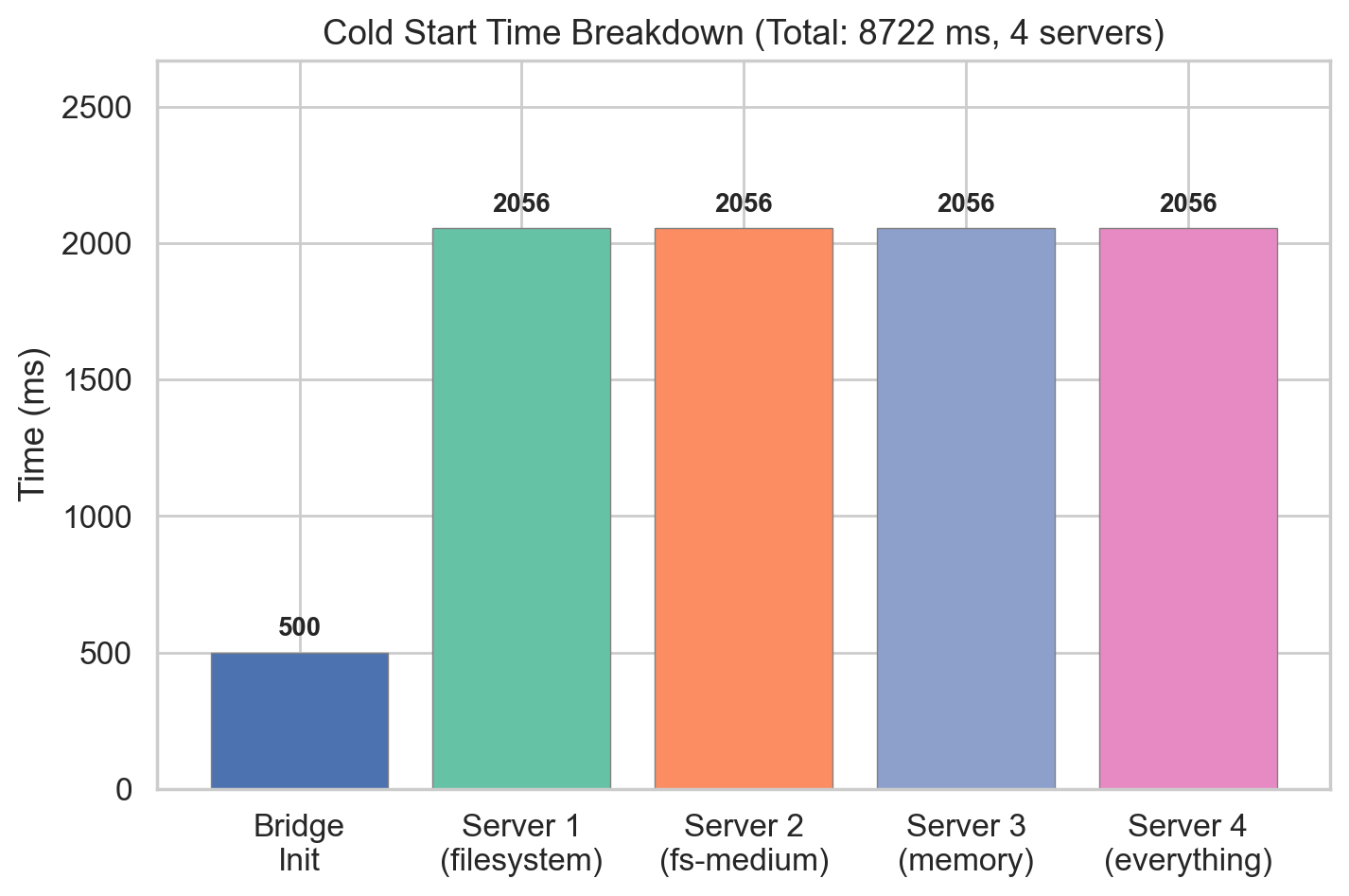}
    \caption{Cold-start time breakdown. Total cold start for 4 MCP servers is 8,722~ms, dominated by Node.js module loading and server process initialization.}
    \label{fig:supp-cold-start}
    \end{figure}
    \FloatBarrier
    
    \begin{figure}[H]
    \centering
    \includegraphics[width=0.90\textwidth]{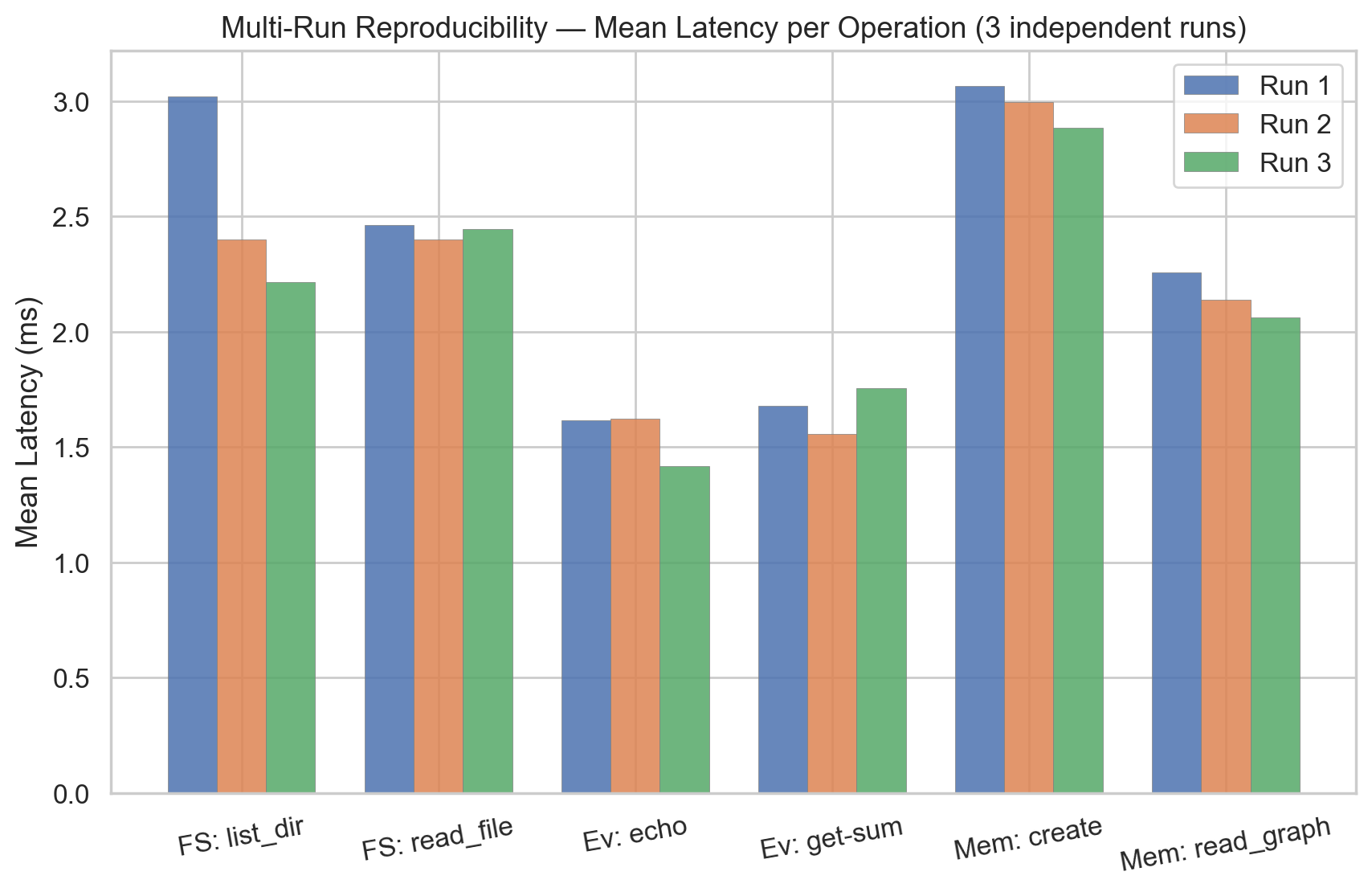}
    \caption{Multi-run reproducibility: mean latency per operation across 3 independent runs. Cross-run variance is low (std $<$ 0.35~ms for all operations), confirming measurement stability.}
    \label{fig:supp-multirun}
    \end{figure}
    \FloatBarrier
    
    \begin{figure}[H]
    \centering
    \includegraphics[width=0.90\textwidth]{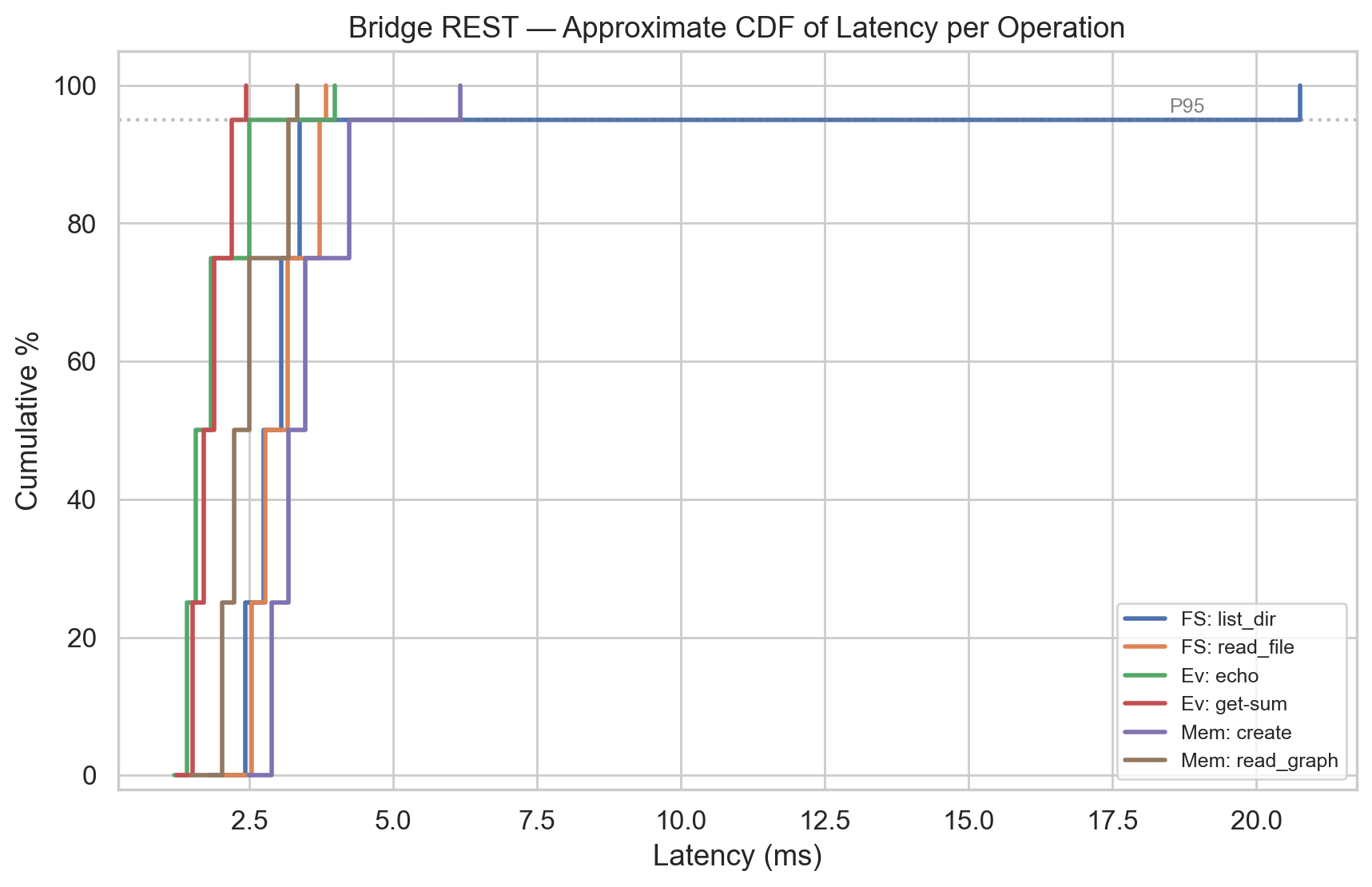}
    \caption{Approximate cumulative distribution function (CDF) of latency per operation. Over 95\% of requests complete within 3.5~ms for all operations.}
    \label{fig:supp-cdf}
    \end{figure}
    \FloatBarrier
    
    \begin{figure}[H]
    \centering
    \includegraphics[width=0.90\textwidth]{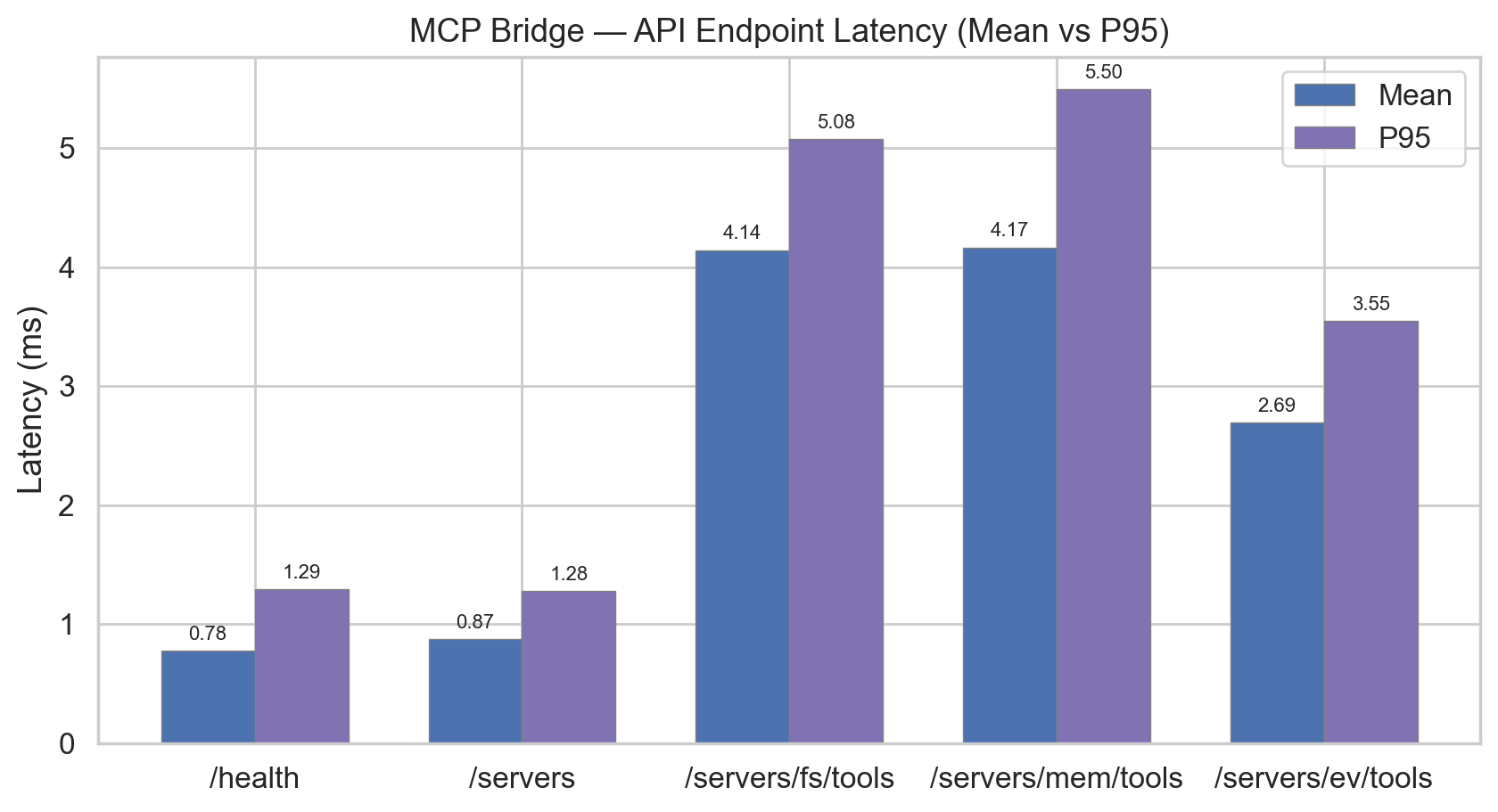}
    \caption{API endpoint latency (mean and P95) for management endpoints (\texttt{/health}, \texttt{/servers}) and tool-listing endpoints. Management endpoints complete in under 1~ms.}
    \label{fig:supp-api-endpoints}
    \end{figure}
    \FloatBarrier
    
    \end{appendices}

\end{document}